\UseRawInputEncoding

\documentclass[aps,prx,twocolumn,showpacs,superscriptaddress]{revtex4-1}

\usepackage[final]{graphicx}
\usepackage{amsmath}
\usepackage{verbatim}
\usepackage{color}
\usepackage{ulem}

\usepackage{siunitx}
\usepackage{threeparttable}

\usepackage{multirow}
\usepackage[colorlinks,
linkcolor=blue,
citecolor=blue,
urlcolor=blue]{hyperref}

\begin{document}
	
	\preprint{}
	
	\title{Suprathermal Electrons from the Anti-Stokes Langmuir Decay Instability Cascade}
	
	%\author{}
	%\email[]{chengzhuo@pku.edu.cn}
	%\thanks{}
	\author{Q. S. Feng} 
	\affiliation{Department of Physics, Atomic and Laser Physics sub-Department, University of Oxford, Clarendon Laboratory, Parks Road, Oxford OX1 3PU, United Kingdom}
	
	\author{R. Aboushelbaya} 
	\affiliation{Department of Physics, Atomic and Laser Physics sub-Department, University of Oxford, Clarendon Laboratory, Parks Road, Oxford OX1 3PU, United Kingdom}
	
	\author{M. W. Mayr} 
	\affiliation{Department of Physics, Atomic and Laser Physics sub-Department, University of Oxford, Clarendon Laboratory, Parks Road, Oxford OX1 3PU, United Kingdom}
	
	\author{B. T. Spiers} 
	\affiliation{Department of Physics, Atomic and Laser Physics sub-Department, University of Oxford, Clarendon Laboratory, Parks Road, Oxford OX1 3PU, United Kingdom}
	
	\author{R. W. Paddock} 
	\affiliation{Department of Physics, Atomic and Laser Physics sub-Department, University of Oxford, Clarendon Laboratory, Parks Road, Oxford OX1 3PU, United Kingdom}
	
	\author{I. Ouatu} 
	\affiliation{Department of Physics, Atomic and Laser Physics sub-Department, University of Oxford, Clarendon Laboratory, Parks Road, Oxford OX1 3PU, United Kingdom}

	\author{R. Timmis} 
	\affiliation{Department of Physics, Atomic and Laser Physics sub-Department, University of Oxford, Clarendon Laboratory, Parks Road, Oxford OX1 3PU, United Kingdom}
	
	\author{R. H. W. Wang} 
	\affiliation{Department of Physics, Atomic and Laser Physics sub-Department, University of Oxford, Clarendon Laboratory, Parks Road, Oxford OX1 3PU, United Kingdom}
	
	\author{L. H. Cao} 
	\email{cao\_lihua@iapcm.ac.cn}
	\affiliation{Institute of Applied Physics and Computational
		Mathematics, Beijing 100094, China}
	\affiliation{HEDPS, Center for
		Applied Physics and Technology, Peking University, Beijing 100871, China}
	
	\author{Z. J. Liu} 
	\affiliation{Institute of Applied Physics and Computational
		Mathematics, Beijing 100094, China}
	\affiliation{HEDPS, Center for
		Applied Physics and Technology, Peking University, Beijing 100871, China}
	
	\author{C. Y. Zheng} 
	\affiliation{Institute of Applied Physics and Computational
		Mathematics, Beijing 100094, China}
	\affiliation{HEDPS, Center for
		Applied Physics and Technology, Peking University, Beijing 100871, China}

	\author{X. T. He} 
	\affiliation{Institute of Applied Physics and Computational
		Mathematics, Beijing 100094, China}
	\affiliation{HEDPS, Center for
		Applied Physics and Technology, Peking University, Beijing 100871, China}
	
	\author{P. A. Norreys} 
	\email{peter.norreys@physics.ox.ac.uk}
	\affiliation{Department of Physics, Atomic and Laser Physics sub-Department, University of Oxford, Clarendon Laboratory, Parks Road, Oxford OX1 3PU, United Kingdom}
	\affiliation{John Adams Institute, Denys Wilkinson Building, Oxford OX1 3RH, United Kingdom}

	%\address[address1]{HEDPS, Center for
	%	Applied Physics and Technology, Peking University, Beijing 100871, China}
	%\address[address2]{Institute of Applied Physics and Computational
	%	Mathematics, Beijing, 100094, China}
	%\address[address3]{Collaborative Innovation Center of IFSA (CICIFSA) , Shanghai Jiao Tong University, Shanghai, 200240, China}
	
	\date{\today}
	
	\begin{abstract}
		The study of parametric instabilities has played a crucial role in understanding energy transfer to plasma and, with that, the development of key applications such as inertial confinement fusion. When the densities are between $0.108n_c\lesssim n_e\lesssim 0.138n_c$ and the electron temperature is $T_e=\SI{2.5}{keV}$, anomalous hot electrons with kinetic energies above \SI{100}{keV} are  generated. Here, a new  electron acceleration mechanism - the anti-Stokes Langmuir decay instability cascade of forward stimulated Raman scattering - is investigated. This mechanism not only explains anomalous energetic electron generation in indirectly driven inertial confinement fusion experiments (and, with that, future mitigation strategies for experiments on the National Ignition Facility), it also provides a new way of accelerating electrons to higher energy for applications such as novel X-ray sources.
		
	\end{abstract}
	
	\maketitle

	% References should be done using the \cite, \ref, and \label 
	\section{Introduction}
	Stimulated Raman scattering (SRS) \cite{LiuCS_1974POF}
	is a three-wave interaction process where an incident light decays into a forward-propagating Langmuir wave and either a backward-scattered (backward-SRS) or a forward-scattered (forward-SRS) electromagnetic wave. Backward-SRS leads to a large energy loss of the incident laser energy in indirect-drive \cite{Glenzer_2007Nature,Glenzer_2010Science} and hybrid-drive inertial confinement fusion (ICF) \cite{He_2016POP,Zhang_2020NP}. The hot electrons generated by the trapping and breaking of SRS-induced Langmuir waves preheat the fusion fuel, while energy losses on the inner cones of beams on the National Ignition Facility have a detrimental effect on the symmetrical compression of fusion capsule, albeit partially compensated by the cross-beam energy transfer process. For these reasons, SRS and hot electron generation should be suppressed to as low a level as possible for indirect-drive. 
	
	A recent review of indirect drive experiments on the National Ignition Facility (NIF) indicated there was a systematic reduction in fuel areal density when compared to those simulated using state of the art radiation hydrodynamic codes \cite{Hinkel_2020HEDP}. In addition, a new hard X-ray imaging device (the eHXI developed by Doeppner et al. \cite{Doppner_2016JI}) was commissioned in 2016. The eHXI instrument is an absolutely-calibrated, time-integrated x-ray imager
	that has the same line-of-sight as the multi-channel, spatially integrating hard x-ray detector FFLEX
	\cite{McDonald_2004RSI} so that it has a side view of indirect-drive inertial
	confinement fusion (ICF) implosion targets. Measurements taken with this instrument have confirmed that electrons with energy above $\SI{100}{keV}$ in the laser irradiated holhraum target make an important contribution to the background noise in the detector, over and above those from the 14.1 MeV fusion neutrons (particularly given the recent remarkable 1.3 MJ fusion yield output result \cite{NIF_2021news}).
	
	The processes responsible for this suprathermal electron generation are still not completely understood. Possible mechanisms that have been studied to date include SRS rescatterings, corresponding Langmuir decay instability (LDI) \cite{Hinkel_2004POP,Langdon_2002PRL} and the two-plasmon decay (TPD) instability \cite{Yan_2012PRL,Pan_2018NF}. 
	Other authors \cite{Bertrand_1995POP,Rousseaux_1992POF,Estabrook_1980PRL,Estabrook_1983POF} have investigated the electrons accelerated by the forward-SRS induced Langmuir waves. 
	
	We draw the attention of the reader to the very elegant work of Winjum {\it et al}. \cite{Winjum_2013PRL} who showed that \SI{100}{keV} electrons are generated through SRS rescattering processes, such as backward-SRS of backward-SRS, backward-SRS of forward-SRS and the corresponding Langmuir decay instability of the rescattering processes when the electron density $n_e\le0.1n_c$. 
	
	We build upon these fascinating insights by showing here that when $T_e=\SI{2.5}{keV}$, and $n_e\gtrsim0.1n_c$, backward-SRS of backward-SRS is not allowed. Similarly, for $n_e\gtrsim0.108n_c$, backward-SRS of forward-SRS is forbidden, since the three-wave matching conditions of rescatterings are not satisfied. Instead, in the region of $n_e\gtrsim0.108n_c$, conclusive evidence is provided, for the first time, that the anomalous hot electrons with energy above \SI{100}{keV} arise from anti-Stokes Langmuir decay instability (denoted as anti-Stokes LDI, or ALDI) cascade of forward-SRS. 
	
\begin{figure*}[!tp]
		\includegraphics[width=2\columnwidth]{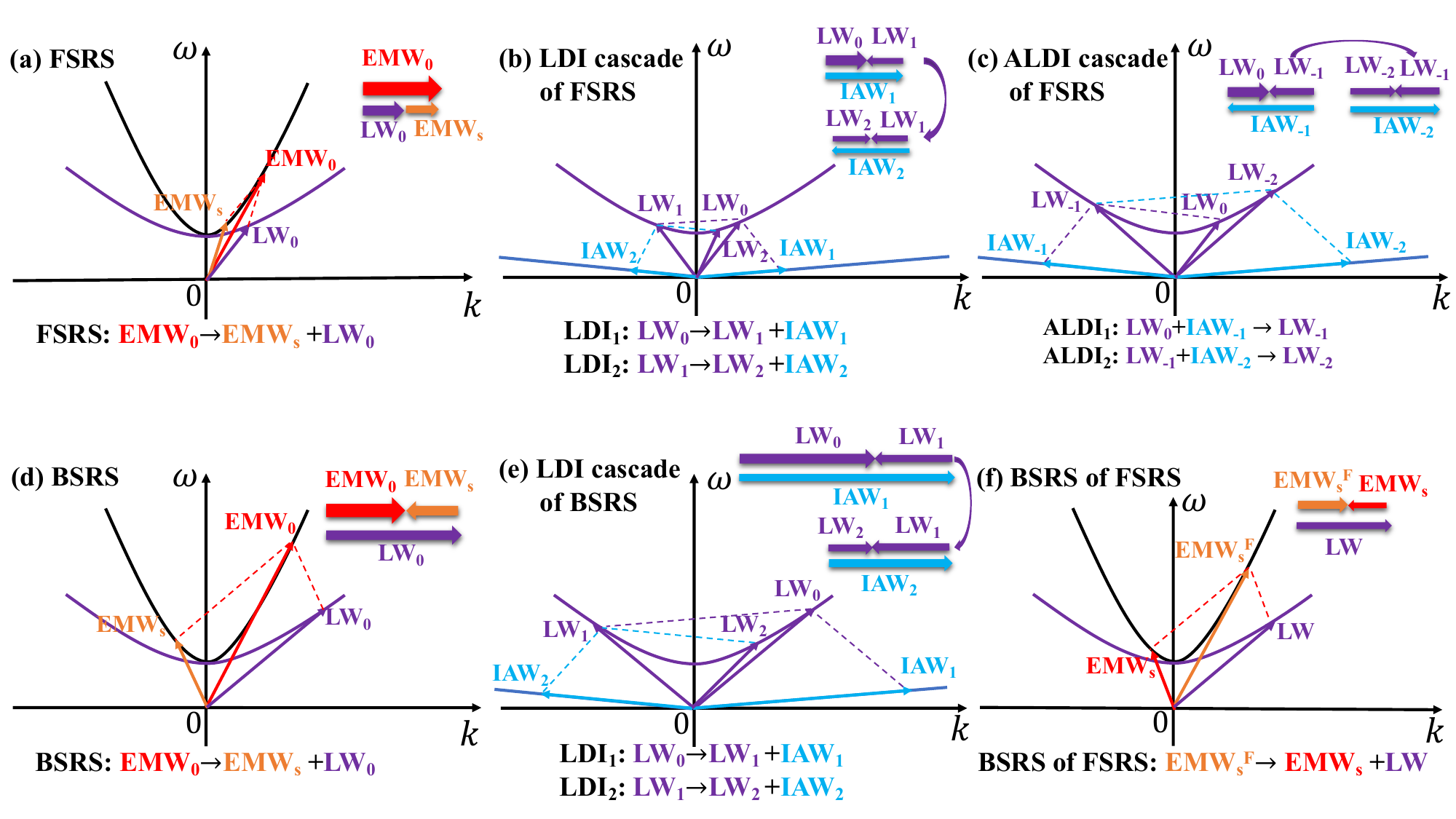}
		\caption{\label{Fig:Schematic}(Color online) Schematic diagram of (a) forward-SRS (FSRS), (b) Langmuir decay instability (LDI) cascade of forward-SRS, (c) anti-Stokes Langmuir decay instability (ALDI) cascade of forward-SRS, (d) backward-SRS (BSRS), (e) Langmuir decay instability cascade of backward-SRS, and (f) backward-SRS of forward-SRS. For completeness, EMW is the electromagnetic wave, LW is the Langmuir wave and IAW is the ion acoustic wave.
		}
	\end{figure*}

	The various wave-wave processes discussed in this article are quite numerous and we have therefore tried to avoid the use of acronyms in the text wherever possible. For clarity, a schematic of the dispersion relationships of the wave-wave processes discussed  in this paper is provided in Figure \ref{Fig:Schematic}.
	
	In indirect drive \cite{Glenzer_2007Nature,Glenzer_2010Science,Kirkwood_2013PPCF} or hybrid drive inertial confinement fusion \cite{He_2016POP,Zhang_2020NP}, suprathermal electrons preheat the fusion fuel, either by direct propagation into the fuel capsule itself or by the secondary generation of hard X-rays that arise when hot electrons propagate into the hohlraum walls or surrounding plasma \cite{Hinkel_2020HEDP,Doppner_2016JI}. On the other hand, in fast ignition \cite{Tabak_1994POP}, the fusion fuel is ignited by collisional stopping of suprathermal electrons that propagate from the critical density to the high-density hot spot on the side of the compressed fusion fuel. Therefore, suprathermal electrons generated by the two stage acceleration mechanism of backward-SRS and forward-SRS in this article might have advantageous effects for inertial fusion by enabling fast ignition using plasma optics and beam-combiners \cite{Kirkwood_2018NP,Kirkwood_2018POP} at lower intensities than previously thought possible. Similarly, the novel mechanism of electron acceleration by anti-Stokes and Stokes Langmuir decay instability cascade of forward-SRS might be a promising mechanism to generate higher energy electrons for radiography purposes \cite{Makse_1995nature,Lemos_2019POP}, complementing those generated with the use of petawatt-class laser pulses.
	
	In this work, we first set the stage by elucidating the parameter regime for the convective and absolute instability of backward-SRS and forward-SRS that is provided by linear theory. Following that, the relevant wave-breaking and particle trapping model is discussed. This model describes the two-stage electron acceleration by backward-SRS and forward-SRS. It is then shown that electrons trapped by the backward-SRS induced Langmuir wave are also trapped and accelerated by the forward-SRS induced Langmuir wave directly in the higher electron density regions where no rescatterings exist as intermediate processes.  
	
	We concentrate on three regions. When $n_e\lesssim 0.108n_c$ (Region I), backward-SRS, backward-SRS of forward-SRS and Langmuir decay instability accelerate electrons to high energy. Only when $0.108n_c\lesssim n_e\lesssim 0.138n_c$ (Region II), are electrons prevented from being accelerated by rescattering and by forward-SRS. However, in Region II, anomalous hot electrons with energies above \SI{100}{keV} are also generated, which is due to the Langmuir wave having a higher phase velocity than the backward-SRS induced Langmuir wave. It is demonstrated that this higher phase velocity Langmuir wave is due to the generation of anti-Stokes Langmuir decay instability in a cascade associated with forward-SRS. When  $n_e\gtrsim0.138n_c$ (Region III), the maximum electron momenta of electrons accelerated by the backward-SRS induced Langmuir wave are much larger than the minimum electron momenta of electrons trapped by the forward-SRS induced Langmuir wave. In this way, the two-stage electron acceleration process by backward-SRS and forward-SRS is allowed. The electron temperature has little effect on the electron acceleration mechanisms, while the decisive factor is the electron density. Lastly, an optimum electron density region with $0.108n_c\lesssim n_e\lesssim 0.138n_c$ is shown to reduce the suprathermal electron population to a low level.

	This paper is arranged as follows. In Section \ref{section:II}, linear theory is used to describe the convective and absolute instability regions of the backward-SRS and forward-SRS processes. The wave-breaking maximum electric field amplitude, that provide insight into the precise conditions required for the two-stage electron acceleration by both backward-SRS and forward-SRS, is also discussed. In Section \ref{section:simulation}, a full-relativistic kinetic Vlasov-Maxwell code is used to: (a) simulate the electron cascade acceleration by backward-SRS and forward-SRS in Region III; (b) study the anti-Stokes Langmuir decay instability cascade of forward-SRS in Region II; and (c) understand the rescattering processes in Region I. In Section \ref{section:EPOCH}, it is shown that both one-dimensional and three-dimensional fully-relativistic kinetic particle-in-cell simulations give similar results to those obtained from Vlasov-Maxwell simulations, confirming that the results are independent of the chosen computational platform and are therefore robust.  In Section \ref{section:Discussions}, the effects of both the electron temperature and inhomogeneous plasma conditions on electron acceleration are discussed. Section \ref{section:conclusions} then summarises and concludes the paper, along with an outline of future research directions.
	
	\begin{table*}
		\caption{\label{table1} The Langmuir wave wave-number $k_L$, electron momentum $p_{\phi}$ and kinetic energy $E_\phi$ at the Langmuir wave phase velocities of the different instabilities. The minus signs of $k_L$ and $p_\phi$ represent waves propagating in the negative direction. The initial electron temperature is $T_e=\SI{2.5}{keV}$.}
		
		%\begin{threeparttable}
		\begin{ruledtabular}
			
			\begin{tabular}{c|ccc|ccc|cc|cc|cc|cc|cc}
				\hline
				& \multicolumn{3}{c|}{BSRS} &\multicolumn{3}{c|}{FSRS}&\multicolumn{2}{c|}{BSRS of FSRS}&\multicolumn{2}{c|}{LDI1 of BSRS}&\multicolumn{2}{c|}{ALDI1 of FSRS}&\multicolumn{2}{c|}{ALDI2 of FSRS}&\multicolumn{2}{c}{ALDI3 of FSRS}
				\\
				
				$n_e$&$k_L$&$p_{\phi}$&$E_{\phi}$&$k_L$&$p_{\phi}$&$E_{\phi}$&$k_L$&$p_{\phi}$&$k_L$&$p_{\phi}$&$k_L$&$p_{\phi}$&$k_L$&$p_{\phi}$&$k_L$&$p_{\phi}$
				\\
				$[n_c]$&$[\lambda_{De}^{-1}]$&$[m_ec]$&$[keV]$&$[\lambda_{De}^{-1}]$&$[m_ec]$&$[keV]$&$[\lambda_{De}^{-1}]$&$[m_ec]$&$[\lambda_{De}^{-1}]$&$[m_ec]$&$[\lambda_{De}^{-1}]$&$[m_ec]$&$[\lambda_{De}^{-1}]$&\multicolumn{1}{c|}{$[m_ec]$}&$[\lambda_{De}^{-1}]$&\multicolumn{1}{c}{$[m_ec]$}
				\\
				\hline

				$0.1$ & $0.33$ & $0.25$ & $16$
				& $0.076$ & $2.4$ & $818$
				& $0.167$ & $0.48$ 
				& $\backslash$ & $\backslash$ 
				& $-0.092$ & $-1.21$ 
				& $0.108$ & $0.88$ 
				& $-0.123$ & $-0.71$ \\
				
				$0.12$ & $0.29$ & $0.28$ & $20$
				& $0.078$ & $2.09$ & $672$
				& $\backslash$ & $\backslash$ 
				& $\backslash$ & $\backslash$ 
				& $-0.094$ & $-1.15$ 
				& $0.109$ & $0.86$ 
				& $-0.125$ & $-0.70$ \\
				
				$0.2$ & $0.18$ & $0.43$ & $46$
				& $0.091$ & $1.25$ & $309$
				& $\backslash$ & $\backslash$ 
				& $-0.169$ & $-0.48$ 
				& $-0.106$ & $-0.90$ 
				& $0.122$ & $0.73$ 
				& $-0.137$ & $-0.61$ 
				\\
				\hline
			\end{tabular}
			
		\end{ruledtabular}
		
		%\end{threeparttable}%
	\end{table*}

	\section{\label{section:II}Theoretical analysis}
	All of the instabilities (backward-SRS, forward-SRS, the various rescattering processes, as well as  the Stokes/anti-Stokes Langmuir decay instability cascades from backward-SRS/forward-SRS) satisfy the three-wave matching conditions resulting from the laws of energy and momentum conservation. By combining the three-wave matching conditions with the dispersion relations of electromagnetic waves (EMW), Langmuir waves (LW) and ion-acoustic waves (IAW), it is possible to calculate the wave number $k_L$, the electron momentum $p_\phi$ and the kinetic energy $E_\phi$ associated with the Langmuir wave phase velocities of the different instabilities, as shown in Table \ref{table1}.

	When the wave number of the Langmuir wave satisfies $k_L\lambda_{De}<0.29$ ($\lambda_{De}$ is the electron Debye length), both the Langmuir decay instability cascade and the anti-Stokes Langmuir decay instability cascade occur more easily, and so are characterised by being in the non-linear wave-wave regime or fluid regime. On the other hand, when  $k_L\lambda_{De}\ge0.29$,  a frequency-broadened spectrum is observed. This spectrum is associated with electron trapping, which places this in the non-linear wave-particle, or kinetic, regime.     \cite{Kline_2005PRL,Kline_2006POP} 
	
	As shown in Table \ref{table1}, three typical electron densities were used in this paper to understand the mechanisms of electron acceleration. Firstly, when $n_e=0.1n_c$, the wave number of the backward-SRS induced Langmuir wave is $k_L^B\lambda_{De}=0.33\ge0.29$. Thus, the Langmuir decay instability cascade of backward-SRS is not allowed, which is the same as the case when $n_e=0.12n_c$. However, the re-scattering of forward-SRS by backward-SRS is allowed. The wave number of rescattering induced Langmuir wave is $k_L\lambda_{De}=0.167<0.29$. The Langmuir decay instability cascade of this re-scattering is also allowed (not shown in Table \ref{table1}). 
	
	Secondly, when $n_e=0.12n_c$ and $n_e=0.2n_c$, re-scattering of forward-SRS by backward-SRS is not allowed because the three wave matching condition of this re-scattering process is not satisfied. However, the Langmuir decay instability cascade of backward-SRS is allowed when $n_e=0.2n_c$ because the wave number of the backward-SRS induced Langmuir wave is $k_L^B\lambda_{De}=0.18<0.29$. 
	
	Thirdly, when $n_e=0.1, 0.12$ and $0.2n_c$, the wave-number of the forward-SRS induced Langmuir wave is $k_L^F\lambda_{De}=[0.076, 0.078, 0.091]<0.29$.  This means that both the Langmuir decay instability and the anti-Stokes Langmuir decay instability of forward-SRS are allowed. 
	
	From the three-wave equations \cite{Forslund_1975POF,Berger_1998POP,HaoLiang_2012CSB}, the theoretical growth rate of SRS scattered light in homogeneous plasmas is \cite{LiuCS_1974POF,DuBois_1974PRL}
	\begin{equation}
	\label{Eq:growth rate}
	\gamma_{tR}=[\frac{2\gamma_{0R}}{\sqrt{|v_{gs}|v_{gL}}}-(\frac{\nu_s}{|v_{gs}|}+\frac{\nu_L}{v_{gL}})]\cdot\frac{|v_{gs}|v_{gL}}{|v_{gs}|+v_{gL}},
	\end{equation}
	where \begin{equation}
	\gamma_{0R}=\frac{1}{4}\sqrt{\frac{\omega_{pe}^2}{\omega_s\omega_L}}k_La_0
	\end{equation}  
	is the maximum temporal growth rate of SRS \cite{Berger_2015PRE,Lindl_2004POP}. $a_0=eE_0/m_e\omega_0$ and $\omega_{pe}=\sqrt{4\pi n_ee^2/m_e}$ are the electron quiver velocity and the electron plasma frequency, respectively. In addition,
	$\nu_i$, $v_{gi}$, $\omega_i$ are the damping rates, group velocities and frequencies of the SRS scattered light ($i=s$) and the Langmuir wave ($i=L$). 
	The damping rate of the backscattered light $\nu_s$ is negligible since it is much lower than the Landau damping $\nu_L$ of the Langmuir wave, i.e., $\nu_s=0$. When $2\gamma_{0R}/\sqrt{|v_{gs}|v_{gL}}>\nu_{L}/v_{gL}$, i.e., $\gamma_{tR}>0$, SRS is in the absolute instability regime. However, when $2\gamma_{0R}/\sqrt{|v_{gs}|v_{gL}}<\nu_{L}/v_{gL}$, i.e., $\gamma_{tR}<0$, SRS is in the convective instability regime.
	The gain of SRS is
	\begin{equation}
	\label{Eq:G_R}
	G_R=2\frac{\gamma_{0R}^2}{\nu_Lv_{gs}}L=\frac{1}{8}\frac{k_L^2a_0^2\omega_{pe}^2}{\nu_Lv_{gs}\omega_{s}\omega_L}L,
	\end{equation}
	where $L$ is the homogeneous plasma length.
	
\begin{figure}[!tp]
		\includegraphics[width=1\columnwidth]{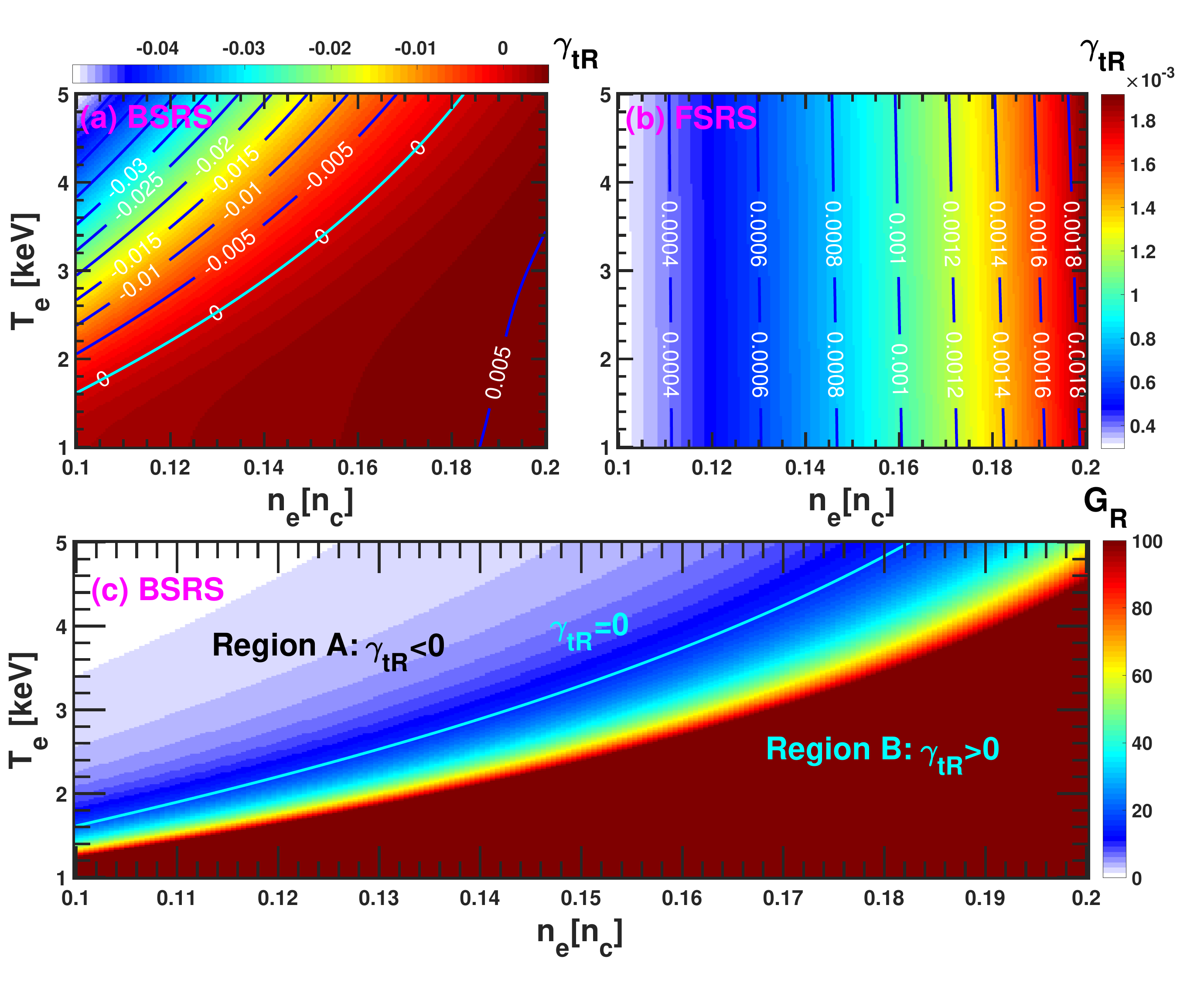}
		\caption{\label{Fig:Convective condition}(Color online) The variation of (a) backward-SRS theoretical growth rate, (b) forward-SRS theoretical growth rate and (c) backward-SRS gain with the electron density and the electron temperature. The incident laser intensity is $I_0=1\times10^{16}\, \SI{}{W/cm^2}$ and the plasma length is $L=400c/\omega_0$.
		}
	\end{figure}

	Fig. \ref{Fig:Convective condition}(a) shows that 
	in the high electron temperature and low electron density region (i.e. where $\gamma_{tR}<0$) the backward-SRS is in the convective instability regime. At the same time, in the low electron temperature and high electron density region (i.e. where $\gamma_{tR}>0$) the backward-SRS is in the absolute instability regime.
	
	Fig. \ref{Fig:Convective condition}(b) also shows the forward-SRS in the parameter region studied in this paper is always in the absolute instability regime due to the low Landau damping of the forward-SRS Langmuir wave. The collision damping of forward-SRS may dominate (which is included in our particle-in-cell simulations in Section \ref{section:EPOCH}). However, the growth rate of the forward-SRS is lower than that of backward-SRS in higher electron density regions.
	
	The contour pictures of backward-SRS gain are shown in Fig. \ref{Fig:Convective condition}(c). Since the wave-number of the backward-SRS induced Langmuir wave is small in the high electron density or low electron temperature region, the Landau damping of backward-SRS induced Langmuir wave is very low. Thus, the backward-SRS gain is very large. When $I_0=1\times10^{16}\, \SI{}{W/cm^2}$, $L=400c/\omega_0$, and $\gamma_{tR}=0$, the backward-SRS gain is $G_R\simeq16$.
	
	Region A in Fig. \ref{Fig:Convective condition}(c) is the convective backward-SRS instability region, where $\gamma_{tR}<0$ and $G_R\lesssim16$. Region B is the absolute backward-SRS instability region, where $\gamma_{tR}>0$ and $G_R\gtrsim16$. If the electron temperature is fixed at $T_e=\SI{2.5}{keV}$, the backward-SRS gain $G_R=3.4$ is very low when $n_e=0.1n_c$, but on the other hand $G_R=2.1\times10^4\rightarrow \infty$ is very high when $n_e=0.2n_c$. When $\gamma_{tR}=0$, the transition between the convective to absolute backward-SRS instability regimes is $n_e\simeq0.128n_c$. If the electron density is fixed at $n_e=0.2n_c$, backward-SRS is an absolute instability and the gain is very high  when the electron temperature lies between $T_e\in[1, 5]$ $\SI{}{keV}$.
	
	The kinetic energy of electrons at the Langmuir wave phase velocity is $E_\phi=1/2m_ev_\phi^2$ and $E_\phi=(\gamma_\phi-1)m_ec^2$ in non-relativistic and relativistic plasmas respectively, where $\gamma_\phi=1/\sqrt{1-(v_\phi/c)^2}$ is the Lorentz factor and $v_\phi$ is the Langmuir wave's phase velocity. This means that when $T_e=\SI{2.5}{keV}$, $ n_e=0.2n_c$, the phase velocity of backward-SRS $v_\phi^B=0.398c$ is on the borderline of the relativistic threshold and the effects of relativity are not immediately obvious. On the other hand, the phase velocity of forward-SRS $v_\phi^F=0.782c$ is much larger than that of backward-SRS, which means that the relativistic effect of electrons with the forward-SRS induced Langmuir wave phase velocity is more obvious. The corresponding kinetic energy of electrons is $E_\phi^B=\SI{46}{keV}$ and  $E_\phi^F=\SI{310}{keV}$ for the backward-SRS and forward-SRS induced Langmuir wave phase velocities, respectively. The maximum or minimum velocity of electrons that are accelerated by the Langmuir wave is related to its phase velocity, i.e., $v_{max}=v_\phi+v_{tr}=v_\phi+2\sqrt{eE_{max}/(m_ek)}$ and $v_{min}=v_\phi-v_{tr}=v_\phi-2\sqrt{eE_{max}/(m_ek)}$. 
	
\begin{figure}[!tp]
		\includegraphics[width=1\columnwidth]{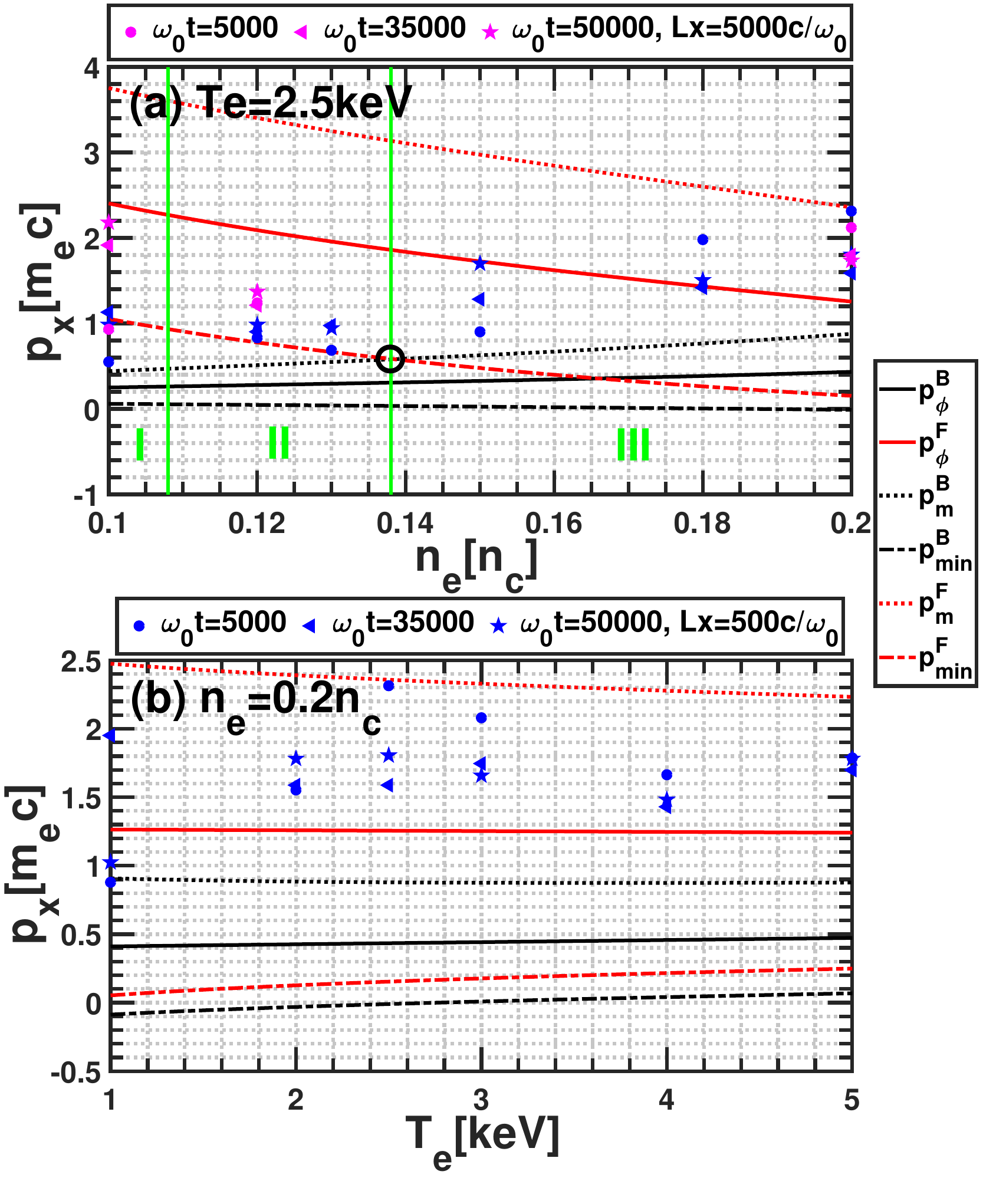}
		\caption{\label{Fig:Theoretical model}(Color online) The variation of $p_x$ with (a) electron density $n_e$ when $T_e=\SI{2.5}{keV}$ and (b) electron temperature $T_e$ when $n_e=0.2n_c$. The Vlasov simulation data points of the maximum electron momenta are chosen at $\omega_0t=5000$ (circle), $\omega_0t=35000$ (triangle) and $\omega_0t=50000$ (pentacle). The simulation data points plotted in pink are in the condition for $I_0=3\times10^{15}\, \SI{}{W/cm^2}$, $L_x=5000c/\omega_0$ (shown in Section \ref{subsection:Vlasov_long_scale}). The simulation data points plotted in blue are in the condition for $I_0=1\times10^{16}\, \SI{}{W/cm^2}$, $L_x=500c/\omega_0$, except that the simulation data points when $T_e=\SI{2.5}{keV}$, $n_e=0.1n_c$ are in the condition for $I_0=5\times10^{16}\, \SI{}{W/cm^2}$ (shown in Section \ref{subsection:Vlasov_short_scale}).
		}
	\end{figure}

	Since the finite electron temperature has to be included explicitly, a warm wave breaking model must be used \cite{Mori_1990PS}. In the limit of slow phase velocity waves, $v_{te}\ll v_{\phi}\ll c$, \cite{Esarey_2009RMP} a warm non-relativistic wave breaking field $E_{max}$ has been calculated using the warm fluid model by Coffey \cite{Coffey_1971POF}. In the ultra-relativistic phase velocity $v_\phi/c=1$ limits, the warm relativistic wave breaking fields given by Katsouleas and Mori \cite{Katsouleas_1988PRL} and by Trines and Norreys \cite{Trines_2006POP} are valid. The warm relativistic wave breaking field is valid for $\gamma_\phi v_{te}\gg1$. In the parameter space of this work, $T_e=\SI{2.5}{keV}$, $n_e\in[0.1, 0.2]n_c$, the phase velocity of the backward-SRS and forward induced Langmuir waves are $v_\phi^B\in[0.24, 0.40]c$ and  $v_\phi^F\in[0.92, 0.78]c$, respectively, while the electron thermal velocity is $v_{te}=0.070c$ for both cases. Therefore, $v_{te}\ll v_\phi^B\ll c$, $v_{te}\ll v_\phi^F<c$ satisfies the warm non-relativistic wave-breaking conditions. Although the phase velocity of the forward-SRS induced Langmuir wave is large, it is also lower than $c$, i.e., $v_\phi^F/c\neq1$. Also $\gamma_\phi\cdot v_{te}/c\in[0.18, 0.11]\ll1$ when $n_e\in[0.1, 0.2]n_c$. For these reasons, the warm fully relativistic wave breaking limits are not valid for the parameter space explored in the current work. Thus, the maximum electric field of the Langmuir wave is calculated by Coffey's warm non-relativistic wave breaking limit \cite{Coffey_1971POF, Mori_1990PS}
	\begin{equation}
	E_{max}=\frac{m_ev_\phi\omega_{pe}}{e}\sqrt{1+2\beta^{1/2}-\frac{8}{3}\beta^{1/4}-\frac{1}{3}\beta},
	\end{equation}
	where $\beta=3v_{te}^2/v_\phi^2$ and $\omega_{pe}, v_{te}$ are the plasma frequency and the electron thermal velocity. In addition, it is possible to calculate the wave number of the Langmuir wave from the Langmuir wave dispersion relation:
	\begin{equation}
	k\simeq\frac{1}{\lambda_{De}}\sqrt{\frac{v_{te}^2}{v_\phi^2-3v_{te}^2}}=\frac{1}{\lambda_{De}}\sqrt{\frac{\beta}{3-3\beta}}.
	\end{equation}
	Thus the maximum momentum is $p_m=\gamma_{max}m_ev_{max}$ and the maximum kinetic energy is $E_m=(\gamma_{max}-1)m_ec^2$. The relationship between the electron momentum, the electron density and electron temperature is shown in Fig. \ref{Fig:Theoretical model}. The electron momenta include those with phase velocities of the backward-SRS induced Langmuir wave ($p_\phi^B$) and the forward-SRS induced Langmuir wave ($p_\phi^F$). Also plotted are the maximum momenta calculated from the wave-breaking maximum electric field of the backward-SRS induced Langmuir wave ($p_m^B$) and the forward-SRS induced Langmuir wave ($p_m^F$), as well as the minimum momenta that are trapped by the backward-SRS induced Langmuir wave ($p_{min}^B$) and the forward-SRS induced Langmuir wave ($p_{min}^F$).

	As shown in Fig. \ref{Fig:Theoretical model}(a), the wave breaking electric field is considered in this model. If the laser intensity is strong enough and the electric field reaches the breaking limit, as in Region III, electrons trapped by the backward-SRS induced Langmuir wave acquire momenta $p_m^B$ larger than the minimum momentum of electrons trapped by the forward-SRS induced Langmuir wave $p_{min}^F$, i.e. $p_m^B>p_{min}^F$. Thus the electrons trapped by the backward-SRS induced Langmuir wave will also be trapped and accelerated by the forward-SRS induced Langmuir wave. 
	
	While in Region II, from the wave-breaking model, no matter how strong the laser intensity is, the electrons trapped by the backward-SRS induced Langmuir wave acquire the maximum momenta $p_m^B$ lower than the minimum momentum associated with being trapped by the forward-SRS induced Langmuir wave $p_{min}^F$, i.e. $p_m^B<p_{min}^F$, even after the wave breaking maximum field is reached. Thus, the two stage electron acceleration by backward-SRS and forward-SRS cannot occur in Region II. 
	In fact, there is another novel mechanism to accelerate the electrons - the anti-Stokes Langmuir decay instability cascade of forward-SRS. This novel mechanism accelerates the electrons to higher momenta than $p_{m}^B$, which will be clarified later. 
	
	In Region I, although there is a large gap between the electron maximum momentum trapped by the backward-SRS induced Langmuir wave $p_m^B$ and the electron minimum momentum which can be trapped by the forward-SRS induced Langmuir wave $p_{min}^F$, the rescattering of SRS and the Langmuir decay instability of the rescattering will accelerate the electrons trapped by the backward-SRS induced Langmuir wave as an intermediate process.
	
	As shown in Fig. \ref{Fig:Theoretical model}(b), when the electron temperature varies, $p_{m}^B$ and $p_{min}^F$ do not vary very much. This illustrates that the electron temperature has little effect on the electron acceleration mechanisms for these conditions.

	\section{\label{section:simulation}Vlasov simulations}
	\subsection{\label{subsection:Vlasov_short_scale}Laser-plasma interactions with small length-scales}

	\begin{figure}[!tp]
		\includegraphics[width=1\columnwidth]{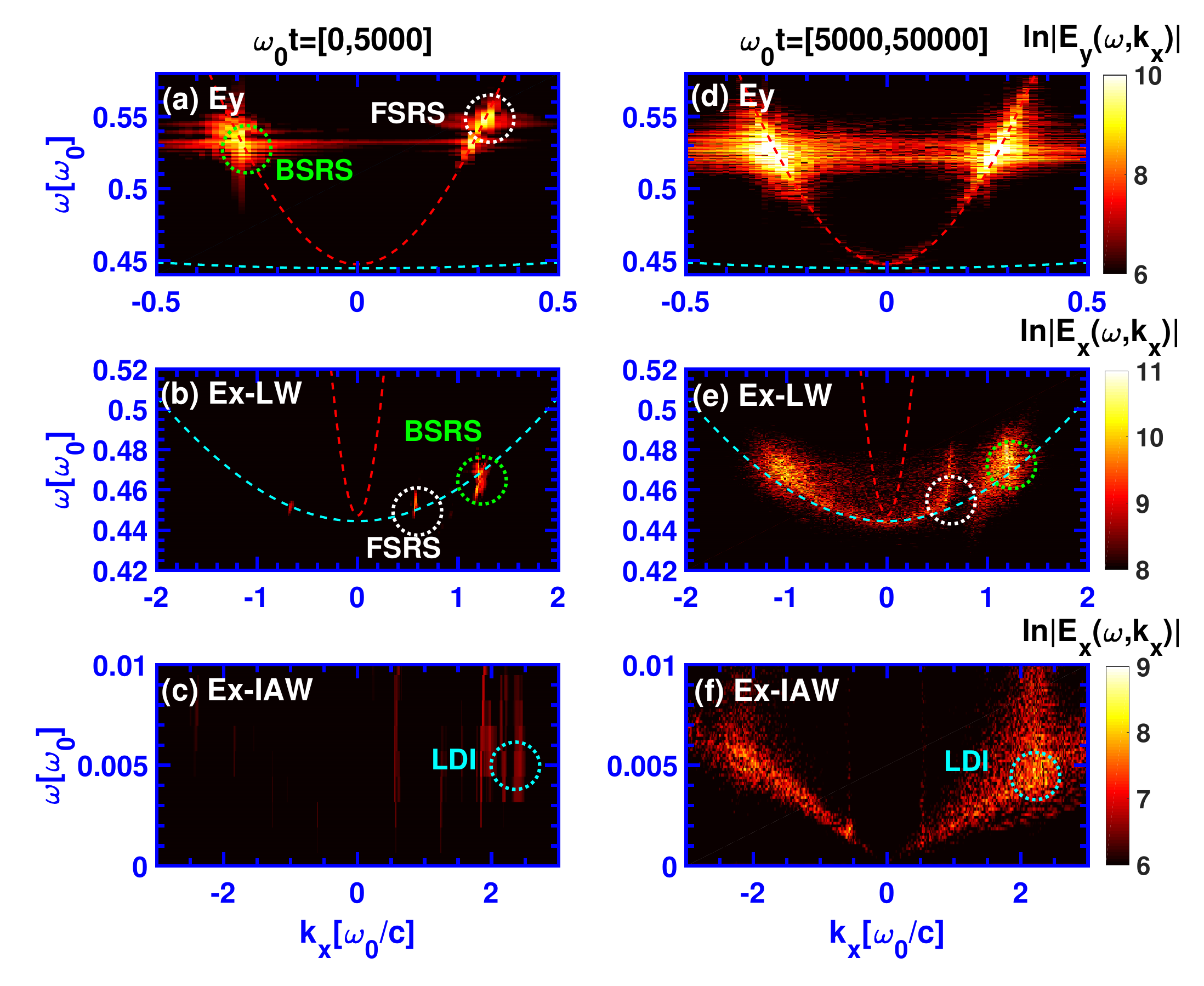}
		\caption{\label{Fig:w_k}(Color online) The dispersion relation of (a), (d) transverse electric field $E_y$ and (b), (c), (e), (f) longitudinal electric field $E_x$. Panels (a)-(c) are during time $t\in[0, 5000]\omega_0^{-1}$ and panels (d)-(e) are during time $t\in[5000, 50000]\omega_0^{-1}$. The parameters are $n_e=0.2n_c, T_e=\SI{2.5}{keV}$, $I_0=1\times10^{16}\, \SI{}{W/cm^2}$ in H plasmas. The cyan dashed line is the weakly relativistic dispersion relation of Langmuir wave \cite{Bergman_2001POP}, i.e. $\omega_L^2=\omega_{pe}^2[1-5/2(v_{te}/c)^2]+3k_L^2v_{te}^2$. }
	\end{figure}
	
	In this Section, we will explore the different mechanisms that accelerate electrons in the three distinct density regions first identified in Fig. \ref{Fig:Theoretical model}(a) and in Table \ref{table1}. In addition, the new mechanism to accelerate electrons - the anti-Stokes Langmuir decay instability cascade of forward-SRS - is explored. 
	
	A one dimensional relativistic kinetic  Vlasov-Maxwell code \cite{Liu_2009POP,Wangq_2018POP,Feng_2017PPCF2,Feng_2018POP,Feng_2020SR} was used to simulate the electron cascade acceleration by the backward-SRS and the forward-SRS induced Langmuir waves. Since the backward-SRS and the forward-SRS induced Langmuir waves propagate in the direction of the incident laser propagation, a one-dimensional simulation is sufficient to study the effects of both processes. A hydrogen (H) plasma was taken as a typical example, as it is commonly used to model the gas fill of  hohlraum targets. The ion temperature was set at $T_i=T_e/3$. The laser was linearly polarized and its wavelength was $\lambda_0=0.351\mu m$. The spatial domain of the simulation was [0, $L_x$] discretized with $N_x=5000$ spatial grid points and spatial step $dx=0.1c/\omega_0$, where $\omega_0$ and $c$ were the frequency and light speed of pump laser in vacuum. Also, the spatial length was $L_x=500c/\omega_0$ with $2\times5\%L_x$ vacuum layers and $2\times5\%L_x$ collision layers on the two sides of the plasma boundaries. These act as strong collision damping layers to damp the electrostatic waves (such as Langmuir waves) at the boundaries. The velocity scale was discretized with $N_v=512$ grid points. The total simulation time was $t_{end}=5\times10^4\omega_0^{-1}$, discretized with a time interval of $dt=0.1\omega_0^{-1}$, to give a total time steps of $N_t=5\times10^5$.
	
	It can be seen from Fig. \ref{Fig:Theoretical model}(a) that when $n_e<0.138n_c$ (for example  $n_e=0.1n_c, 0.12n_c$) and $T_e=\SI{2.5}{keV}$, $p_{min}^F>p_m^B$ and the maximum momenta of the electrons are not larger than $p_\phi^F$. This verifies the fact that the electrons trapped by the backward-SRS induced Langmuir wave cannot be trapped and accelerated directly by the forward-SRS induced Langmuir wave in Regions I and II. However, when $n_e>0.138n_c$ (Region III), $p_{min}^F<p_m^B$, the electrons trapped by the backward-SRS induced Langmuir wave are also trapped and accelerated by the forward-SRS induced Langmuir wave. One sees that the maximum momenta from Vlasov simulations at $n_e=0.18n_c, 0.2n_c$ are larger than the momentum  ($p_\phi^F$) at the phase velocity of the forward-SRS induced Langmuir wave. This points towards the existence of a transitional density $n_e\simeq0.138n_c$ for the electron acceleration cascade when the electron temperature is $T_e=\SI{2.5}{keV}$.
	
	One also sees that when $n_e\lesssim0.108n_c$ (Region I), backward-SRS of forward-SRS occurs, and this accelerates electrons as an intermediate process. However, in Region II ($0.108n_c\lesssim n_e\lesssim 0.138n_c$), there are electrons with momenta larger than $p_m^B$, which cannot have been accelerated solely by the backward-SRS induced Langmuir wave. 
	
	When the electron density is limited to $0.108n_c\lesssim n_e\lesssim0.138n_c$, suprathermal electrons are suppressed. However, it is important to note that when the density is in the range $n_e\gtrsim0.128n_c$,  $\nu_L/(2\gamma_R)\equiv\nu_L/(2\gamma_{0R})\sqrt{v_{gs}/v_{gL}}<1$ and $n_e\lesssim0.128n_c$, $\nu_L/(2\gamma_R)>1$, the backward-SRS process is an absolute and convective instability, respectively. This means that in order to reduce the backward-SRS reflectivity and keep the number of hot electrons to a minimum, the electron density must be limited in $n_e\lesssim0.128n_c$ in order to maintain backward-SRS as a convective instability. Thus, the electron density in the region $0.108n_c\lesssim n_e\lesssim0.128n_c$ (when $T_e=\SI{2.5}{keV}$) is the best choice to control suprathermal electron generation and backward-SRS in indirect drive ICF.  
	
	From Fig. \ref{Fig:Theoretical model}(b), $p_{min}^F<p_m^B$ in the parameter range $T_e\in[1, 5]$ $\SI{}{keV}$. This means that the electrons trapped by backward-SRS are also trapped and accelerated by the forward-SRS provided that the laser intensity is high enough. Comparing Fig. \ref{Fig:Theoretical model}(a) and \ref{Fig:Theoretical model}(b), it is clear that the electron momenta are more sensitive to the electron density than the electron temperature.

	%  \begin{figure}[!tp]
	%  	\includegraphics[width=1\columnwidth]{Fig3.eps}
	%  	
	%  		\caption{\label{Fig:PhasePicture}(Color online) The phase picture of electrons in (a) the total space region and (b) partial space region, and (c) corresponding distribution averaged in space scope $[50, 450]c/\omega_0$ at the time of $\omega_0t=5000$. (d), (e) The phase picture of ions in the total space region and partial space region at the time of $\omega_0t=5\times10^{4}$. (f) The distribution averaged in space at different time. The parameters are $n_e=0.2n_c, T_e=2.5keV, I_0=1\times10^{16}W/cm^2$. 
	%  		}
	%  \end{figure}
	%  
	
	To clarify the physics of the new electron acceleration mechanism in Region II, let us first discuss the two-stage electron acceleration by backward-SRS and forward-SRS in Region III, shown by Figs. \ref{Fig:w_k}-\ref{Fig:f_Ek} when $n_e=0.2n_c$ and $T_e=\SI{2.5}{keV}$. 
	
	\subsubsection{Two-stage electron acceleration by backward-SRS and forward-SRS in Region III}
	Figure \ref{Fig:w_k} demonstrates the spectra of the transverse electric field $E_y$ and the longitudinal electric field $E_x$ at different times. At early times, i.e. $\omega_0t\in[0, 5000]$, backward-SRS and forward-SRS develop from zero to their maximum levels, and the spectra are discrete. At the same time, the Langmuir decay instability of backward-SRS develops but is not immediately obvious at this stage. In the interval $\omega_0t\in[5000, 50000]$, backward-SRS and forward-SRS both saturate, and the spectra of backward-SRS broaden, as shown in Figs. \ref{Fig:w_k}(d) and \ref{Fig:w_k}(e), since the amplitude of the backward-SRS and the forward-SRS scattered electromagnetic waves are strong and the beam acoustic mode (BAM) \cite{Yin_2006POP} develops. The Langmuir decay instability and its second-stage, denoted as LDI2, are now more clearly observed, as shown in Figs. \ref{Fig:w_k}(e) and \ref{Fig:w_k}(f).

	\begin{figure}[!tp]
		%	\centering
		\includegraphics[width=1.08\columnwidth]{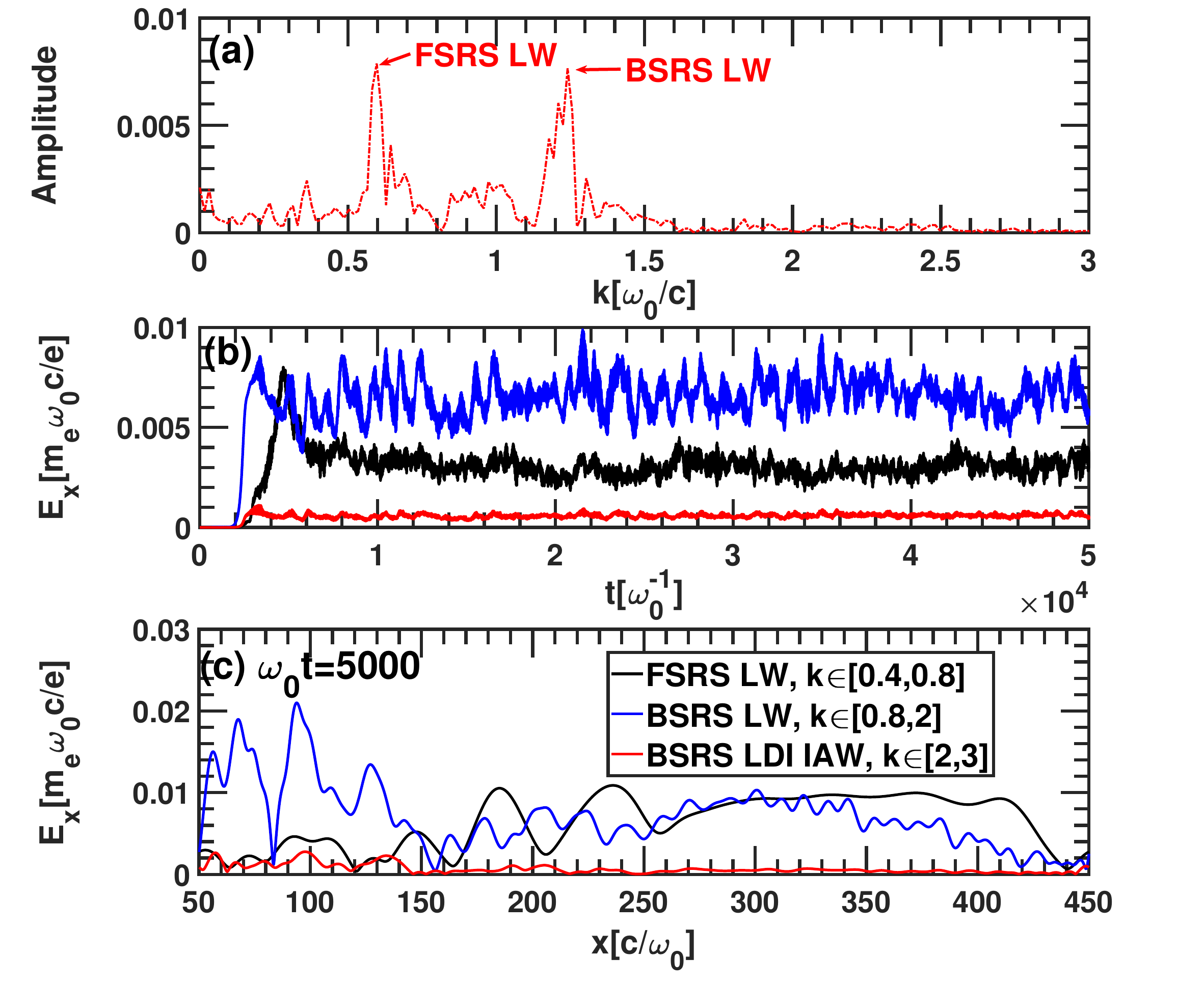}
		\caption{\label{Fig:Ex_t_Ex_x}(Color online) (a) The wave-number spectrum of the longitudinal electric field $E_x$ when $\omega_0t=5000$. (b) The time evolution and (c) spatial evolution of Langmuir wave of forward-SRS ($k\in[0.4, 0.8]\omega_0/c$, black line), Langmuir wave of backward-SRS ($k\in[0.8, 2]\omega_0/c$, blue line) and IAW of LDI of backward-SRS ($k\in[2, 3]\omega_0/c$, red line). The electric field $E_x$ in (b) is averaged in space $[50, 450]c/\omega_0$. The parameters are the same as Fig. \ref{Fig:w_k}.
		}
	\end{figure}
	
	\begin{figure*}
		\begin{minipage}{0.8\textwidth} 
			\includegraphics[width=0.98\columnwidth]{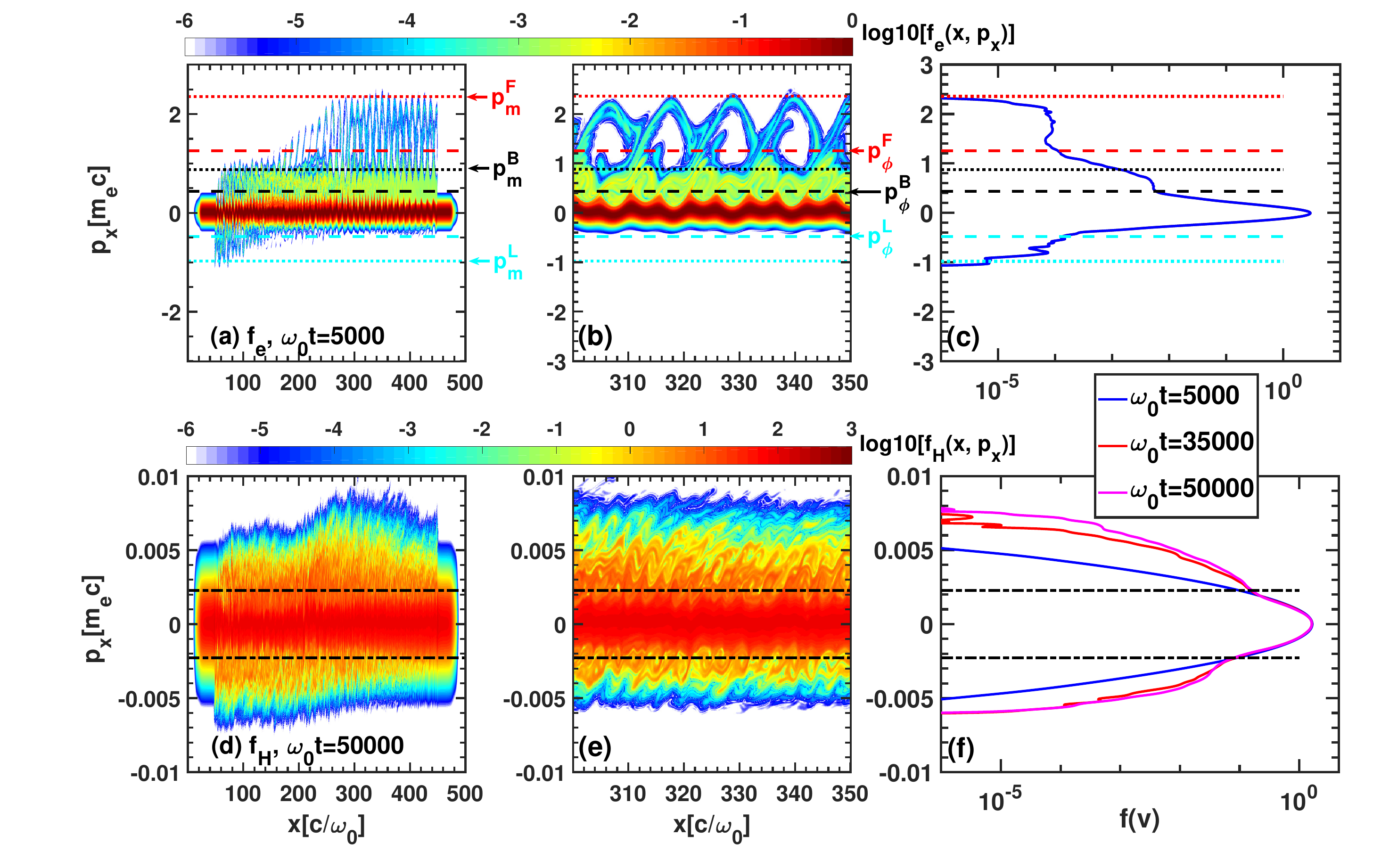}
		\end{minipage}% 
		\begin{minipage}{0.2\textwidth}
			\caption{\label{Fig:PhasePicture}(Color online) The phase space of electrons in (a) the whole space region and (b) partial space region. (c) Corresponding momentum distribution averaged in space $[50, 450]c/\omega_0$ at the time of $\omega_0t=5000$. (d), (e) The phase space of ions in the whole space region and partial space region at the time of $\omega_0t=5\times10^{4}$. (f) The distribution averaged in space at different times. The parameters are $n_e=0.2n_c, T_e=\SI{2.5}{keV}$, $I_0=1\times10^{16}\, \SI{}{W/cm^2}$, the same as in Fig. \ref{Fig:w_k}. 
			}
		\end{minipage}
	\end{figure*}

	\begin{figure}[!tp]
		%	\centering
		\includegraphics[width=1\columnwidth]{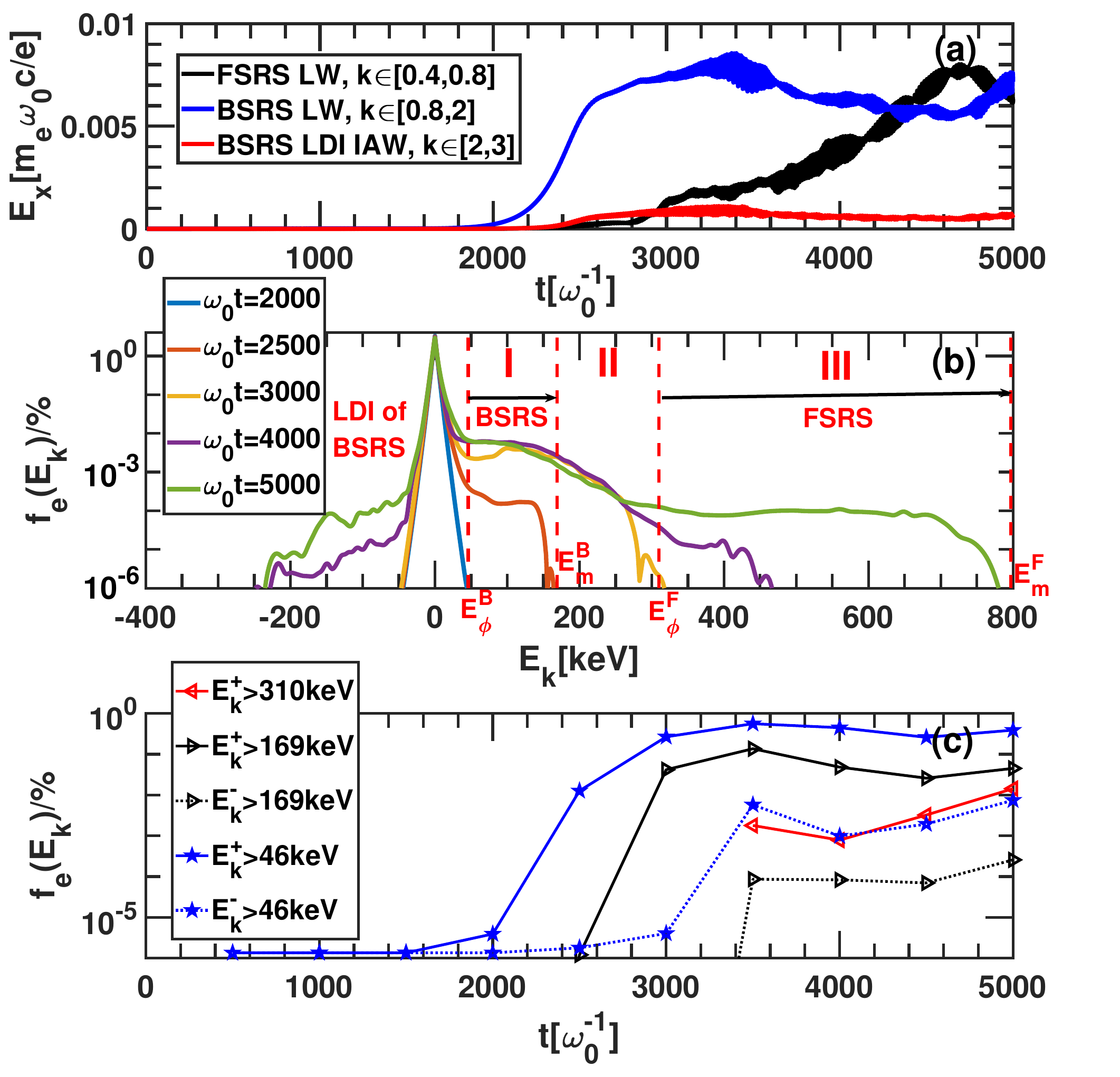}
		\caption{\label{Fig:f_Ek}(Color online) (a) The time evolution of the Langmuir wave of forward-SRS ($k_L^F=0.579\omega_0/c\in[0.4, 0.8]\omega_0/c$), the Langmuir wave of backward-SRS ($k_L^B=1.18\omega_0/c\in[0.8, 2]\omega_0/c$) and IAW of LDI of backward-SRS ($k_A^L=2.26\omega_0/c\in[2, 3]\omega_0/c$). The electric field $E_x$ is averaged in space $[50, 450]c/\omega_0$. (b) The distribution of kinetic energy at different times. (c) The evolution in time of the ratios of the electrons in different energy ranges. $E_k^{\pm}$ represents the kinetic energy of electrons with positive velocity ($+$) and negative velocity ($-$). From Fig. \ref{Fig:Theoretical model}, $E_\phi^B=\SI{46}{keV}$, $E_m^B=\SI{169}{keV}$, $ E_\phi^F=\SI{310}{keV}$, $E_m^F=\SI{797}{keV}$ when $n_e=0.2n_c, T_e=\SI{2.5}{keV}$, $I_0=1\times10^{16}\, \SI{}{W/cm^2}$.
		}
	\end{figure}

	Figure \ref{Fig:Ex_t_Ex_x} shows the wave-number spectrum (Fig. \ref{Fig:Ex_t_Ex_x}(a)), time evolution (Fig. \ref{Fig:Ex_t_Ex_x}(b)) and spatial distribution (Fig. \ref{Fig:Ex_t_Ex_x}(c)) of the electrostatic field $E_x$. From Fig. \ref{Fig:Ex_t_Ex_x}(a), the wave number of the backward-SRS induced Langmuir wave is $k_L^B=1.24\omega_0/c$ and that of the forward-SRS induced Langmuir wave is $k_L^F=0.597\omega_0/c$, which are close to the theoretical values $k_L^B=1.18\omega_0/c$ and $k_L^F=0.579\omega_0/c$.
	
	The theoretical wave-number of the ion acoustic wave  generated by the Langmuir decay instability of backward-SRS is $k_{IAW}=2.260\omega_0/c$, and this is demonstrated in Fig. \ref{Fig:w_k}(f) but is not so obvious in Fig. \ref{Fig:Ex_t_Ex_x}(a) since the ion acoustic wave from the Langmuir decay instability is much weaker than the backward-SRS induced Langmuir wave. Fig. \ref{Fig:Ex_t_Ex_x}(b) shows the time evolution of the Langmuir wave of forward-SRS, the Langmuir wave of backward-SRS and the ion acoustic wave generated by Langmuir decay instability of backward-SRS. 
	
	The backward-SRS induced Langmuir wave increases more quickly and  saturates at earlier times than the forward-SRS induced Langmuir wave. The theoretical growth rates  \cite{Forslund_1975POF} of backward-SRS and forward-SRS are  $\gamma_{tR}^B=5.2\times10^{-3}\omega_0$ and $\gamma_{tR}^F=1.9\times10^{-3}\omega_0$, respectively. From this, one sees that the Langmuir wave of backward-SRS increases more quickly than the Langmuir wave of forward-SRS, since $\gamma_{tR}^B>\gamma_{tR}^F$. 
	
	Fig. \ref{Fig:Ex_t_Ex_x}(c) shows the spatial distribution of the forward-SRS induced Langmuir wave, the backward-SRS induced Langmuir wave and the ion acoustic wave generated by Langmuir decay instability of backward-SRS. Backward-SRS is distributed in the entire simulation space. Among $x\in[50, 150]c/\omega_0$, backward-SRS is much stronger than the forward-SRS, thus one concludes that the electrons are only accelerated by the backward-SRS induced Langmuir waves in this space. On the other hand, within the $x\in[150, 450]c/\omega_0$ region, forward-SRS becomes large and coexists with backward-SRS. One therefore concludes that the two-stage acceleration of electrons occurs in this region.

	\begin{figure*}
		\begin{minipage}{0.8\textwidth} 
			\includegraphics[width=0.98\columnwidth]{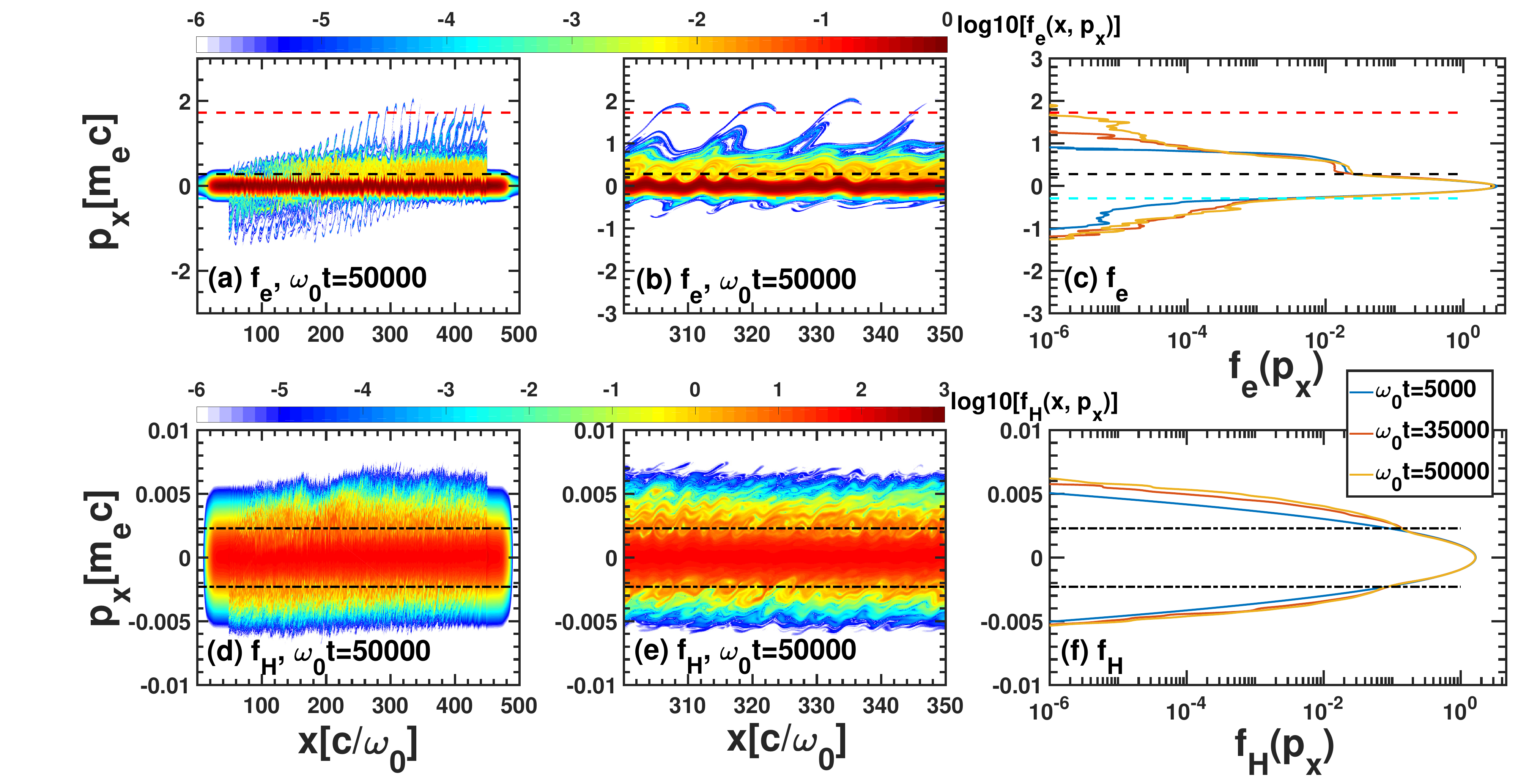}
		\end{minipage}% 
		\begin{minipage}{0.2\textwidth}
			\caption{\label{Fig:PhasePicture_ne=0.15}(Color online) The phase space of electrons in (a) the total space region and (b) partial space region, and (c) corresponding distribution averaged in space range $[50, 450]c/\omega_0$. (d), (e) The phase space of ions in the total space region and partial space region at the time of $\omega_0t=5\times10^{4}$. (f) The distribution averaged in space at different time. The parameters are $n_e=0.15n_c, T_e=\SI{2.5}{keV}$, $ I_0=1\times10^{16}\, \SI{}{W/cm^2}$. 
			}
		\end{minipage}
	\end{figure*}
	
	Figure \ref{Fig:PhasePicture}(a) shows the electron distribution in the total simulation phase space. The electrons between $x\sim[50, 150]c/\omega_0$ are only accelerated by the backward-SRS induced Langmuir wave. Also, the electrons accelerated by the backward-SRS induced Langmuir wave are trapped and gradually accelerated by the forward-SRS induced Langmuir wave in $x\sim[150, 250]c/\omega_0$, since the electric field of the forward-SRS induced Langmuir wave gradually increases in this region. Also, when $x\in[250, 450]c/\omega_0$, the electrons are accelerated to nearly their maximum momenta, close to the theoretical values $p_m^F=2.4m_ec$. 
	
	Fig. \ref{Fig:PhasePicture}(b) presents a clearer demonstration of the two-stage acceleration of electrons by backward-SRS and forward-SRS in Region II ($n_e=0.2n_c$). The electrons trapped by the backward-SRS induced Langmuir wave are accelerated to the maximum momenta $p_m^B=0.88m_ec$, which are then further trapped by the forward-SRS induced Langmuir wave and accelerated to the maximum momenta $p_m^F=2.4m_ec$. Since $p_m^B>p_{min}^F$, the electrons with momenta $p\in[p_{min}^F, p_m^B]$ are trapped by both the forward-SRS and the backward-SRS induced Langmuir waves, this leads to the distortion of the phase-space distribution of electrons with momenta $p\in[p_{min}^F, p_m^B]$. 
	
	The momentum distribution of electrons averaged in the simulation space is shown in Fig. \ref{Fig:PhasePicture}(c). This shows that the distribution of electrons is flattened around the phase velocity of the backward-SRS induced Langmuir wave, meaning that the number of electrons accelerated by the backward-SRS induced Langmuir wave increases. When the maximum momenta of hot electrons accelerated by the backward-SRS induced Langmuir wave are higher than the minimum momenta of electrons that are trapped by the forward-SRS induced Langmuir wave, i.e., $p_m^B>p_{min}^F$, the hot electrons are further trapped and accelerated by the forward-SRS induced Langmuir wave. 
	
	When the incident laser intensity is high enough, the Langmuir waves generated by backward-SRS and forward-SRS will be large enough to reach the maximum amplitude of electric field given by the wave breaking limit discussed in Section \ref{section:II}. In our simulations with $I_0=1\times10^{16}\, \SI{}{W/cm^2}$, the maximum momentum of electrons generated by the two-stage acceleration of backward-SRS and forward-SRS is $p{'}_m^F=2.3m_ec$, which is slightly lower than the theoretical value $p_m^F=2.4m_ec$, since the maximum electric field in the simulation is lower than the theoretical wave breaking value. 
	
	Figs. \ref{Fig:PhasePicture}(d)-(f) demonstrate the distribution of H ions. Since $n_e=0.2n_c$, the Langmuir decay instability threshold \cite{Karttunen_1981PRA} is very low, so the Langmuir decay instability of backward-SRS and the associated cascade \cite{Feng_2018POP} occur quite easily. Due to the combination of Langmuir decay instabilities (labelled LDI1 and LDI2 in Fig. \ref{Fig:Schematic}), the positive directed ion acoustic wave (generated by the first Langmuir decay instability and the opposite directed ion acoustic wave by the second) will trap the H ions, as shown in Figs. \ref{Fig:PhasePicture}(d)-(f). It should be noted that the Langmuir decay instability produces a negative-propagation direction Langmuir wave which traps electrons, as shown in Fig. \ref{Fig:PhasePicture}(a) and Fig. \ref{Fig:PhasePicture}(c). 
	Higher order Langmuir decay instabilities (e.g. those that might be labeled LDI3, following the nomenclature in the caption of Fig. \ref{Fig:Schematic}) generate a Langmuir wave with higher phase velocity, thus the maximum momentum of negative directed electrons is larger than the theoretical value $p_m^L$ which considers only the first-stage of the Langmuir decay instability.

	\begin{figure*}
		\begin{minipage}{0.8\textwidth} 
			\includegraphics[width=0.98\columnwidth]{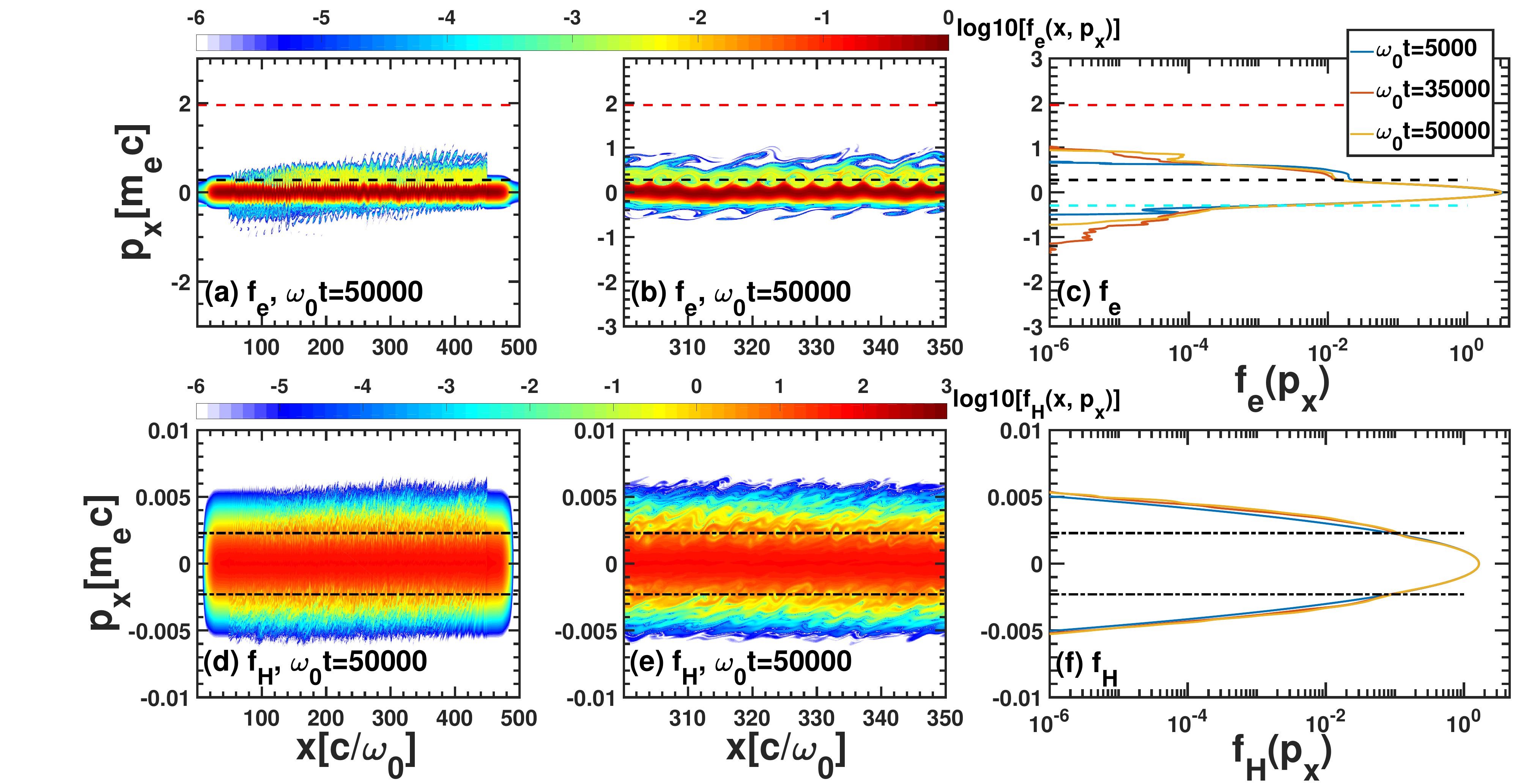}
		\end{minipage}% 
		\begin{minipage}{0.2\textwidth}
			\caption{\label{Fig:PhasePicture_ne=0.13}(Color online) The phase space of electrons in (a) the whole space region and (b) partial space region. (c) Corresponding distribution averaged in space $[50, 450]c/\omega_0$. (d), (e) The phase space of ions in the total space region and partial space region at the time of $\omega_0t=5\times10^{4}$. (f) The distribution averaged in space at different times. The parameters are $n_e=0.13n_c, T_e=\SI{2.5}{keV}$, $ I_0=1\times10^{16}\, \SI{}{W/cm^2}$. 
			}
		\end{minipage}
	\end{figure*}
	
	\begin{figure*}
		\begin{minipage}{0.8\textwidth} 
			\includegraphics[width=0.98\columnwidth]{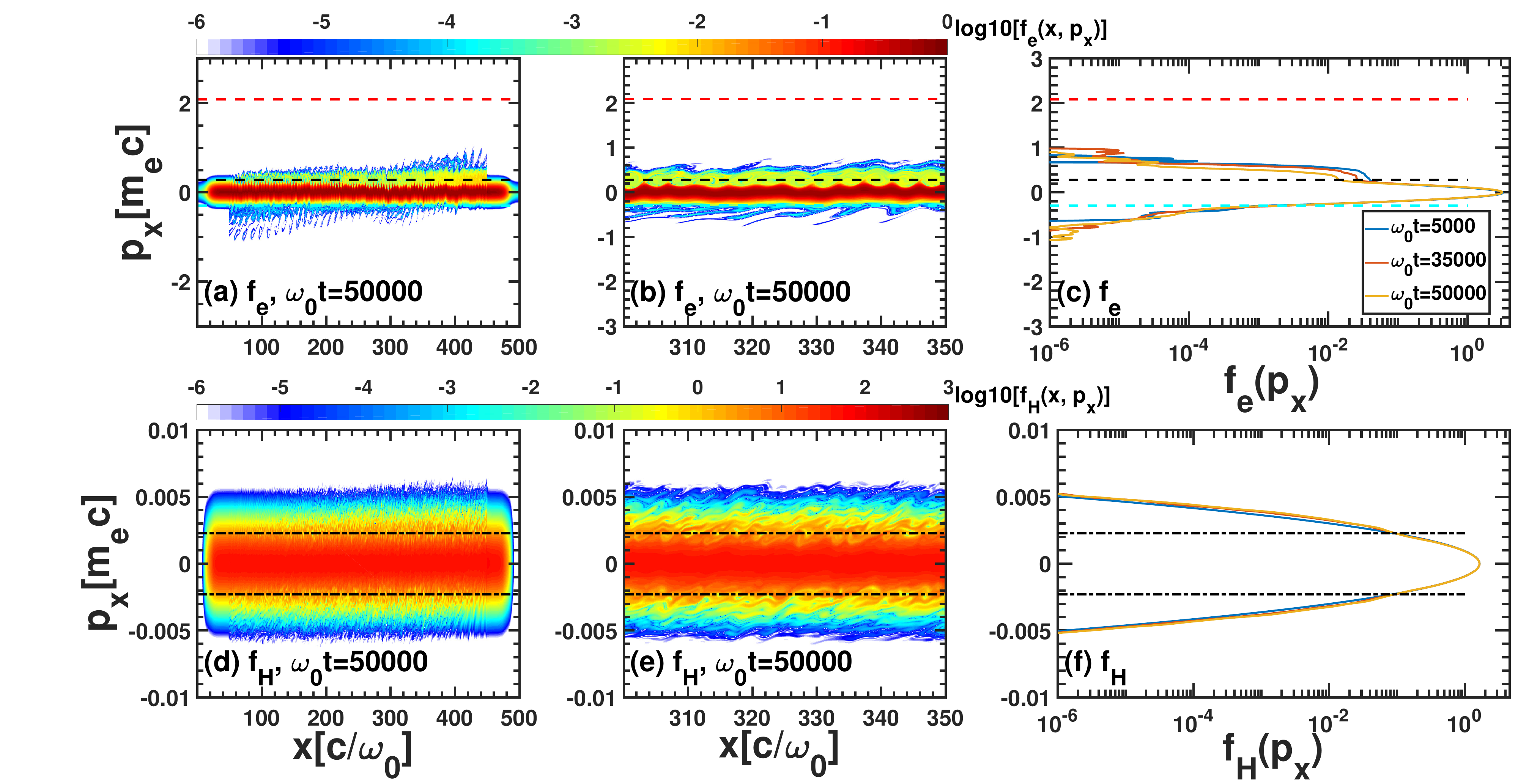}
		\end{minipage}% 
		\begin{minipage}{0.2\textwidth}
			\caption{\label{Fig:PhasePicture_ne=0.12}(Color online) The phase space of electrons in (a) the whole space region and (b) partial space region. (c) Corresponding distribution averaged in space range $[50, 450]c/\omega_0$. (d), (e) The phase space of ions in the total space region and partial space region at the time of $\omega_0t=5\times10^{4}$. (f) The distribution averaged in space at different times. The parameters are $n_e=0.12n_c, T_e=\SI{2.5}{keV}$, $ I_0=1\times10^{16}\, \SI{}{W/cm^2}$. 
			}
		\end{minipage}
	\end{figure*}

	As shown in Fig. \ref{Fig:f_Ek}(a), the backward-SRS induced Langmuir wave develops from $t\simeq2000\omega_0^{-1}$ and saturates at $t\simeq3400\omega_0^{-1}$, while the forward-SRS induced Langmuir wave develops from $t\simeq2500\omega_0^{-1}$ and saturates at $t\simeq4800\omega_0^{-1}$.
	
	Fig. \ref{Fig:f_Ek}(b) shows that, at $\omega_0t=2000$, the distribution of electrons is initially Maxwellian and the electrons are not accelerated. There are three processes associated with the two-stage electron acceleration by backward-SRS and forward-SRS. 
	
	Process I: the backward-SRS induced Langmuir wave only accelerates the electrons from $E_\phi^B=\SI{46}{keV}$ to $E_m^B=\SI{169}{keV}$ during $\omega_0t\simeq[2000, 2500]$.
	
	Process II: the low amplitude forward-SRS induced Langmuir wave traps the electrons that are accelerated by the backward-SRS induced Langmuir wave during $\omega_0t\simeq[2500, 3000]$, the energy gap is from $E_m^B=\SI{169}{keV}$ to $E_\phi^F=\SI{310}{keV}$. 
	
	Process III: the large amplitude forward-SRS induced Langmuir wave traps and accelerates electrons from $E_\phi^F=\SI{310}{keV}$ to nearly $E_m^F=\SI{797}{keV}$ during $\omega_0t\simeq[3000, 5000]$. 
	
	The electron ratio is shown in Fig. \ref{Fig:f_Ek}(c). The ratio of electrons with energy larger than $E_\phi^B=\SI{46}{keV}$ develops from $3.9\times10^{-6}\%$ ($\omega_0t=2000$) to $0.56\%$ ($\omega_0t=3500$). When $\omega_0t>3500$, the ratio of electrons with $E_k>\SI{46}{keV}$ trapped by the backward-SRS induced Langmuir wave decreases due to the saturation and decrease of the backward-SRS induced Langmuir wave amplitude as shown in Fig. \ref{Fig:f_Ek}(a). The ratio of hot electrons with $E_k>E_m^B=\SI{169}{keV}$ develops from $1.2\times10^{-6}\%$ ($\omega_0t=2500$) to $0.14\%$ ($\omega_0t=3500$) and saturates and decreases after $\omega_0t=3500$, which is consistent to the backward-SRS induced Langmuir wave amplitude. This part of electrons with $E_k^+\in[E_m^B, E_\phi^F]$ are trapped by the backward-SRS induced Langmuir wave to reach $E_m^B$ and then trapped by the forward-SRS induced Langmuir wave. As a result, they are related to the backward-SRS induced Langmuir wave amplitude. 
	
	The ratio of electrons with $E_k^+>E_\phi^F=\SI{310}{keV}$ is from $1.8\times10^{-3}\%$ ($\omega_0t=3500$) to $0.014\%$ ($\omega_0t=5000$). These electrons are only accelerated by the large-amplitude forward-SRS induced Langmuir wave. The number of  negative-direction propagating electrons with $E_k^->\SI{46}{keV}$ and $E_k^->\SI{169}{keV}$ is very small as shown in Fig. \ref{Fig:f_Ek}(c), which are generated by the Langmuir decay instability of backward-SRS. Although the number of suprathermal electrons with $E_k^+>E_m^F=\SI{310}{keV}$ is very small, the kinetic energy of these electrons are very high, which preheat the fusion fuel.

	\begin{figure}[!tp]
		\includegraphics[width=1\columnwidth]{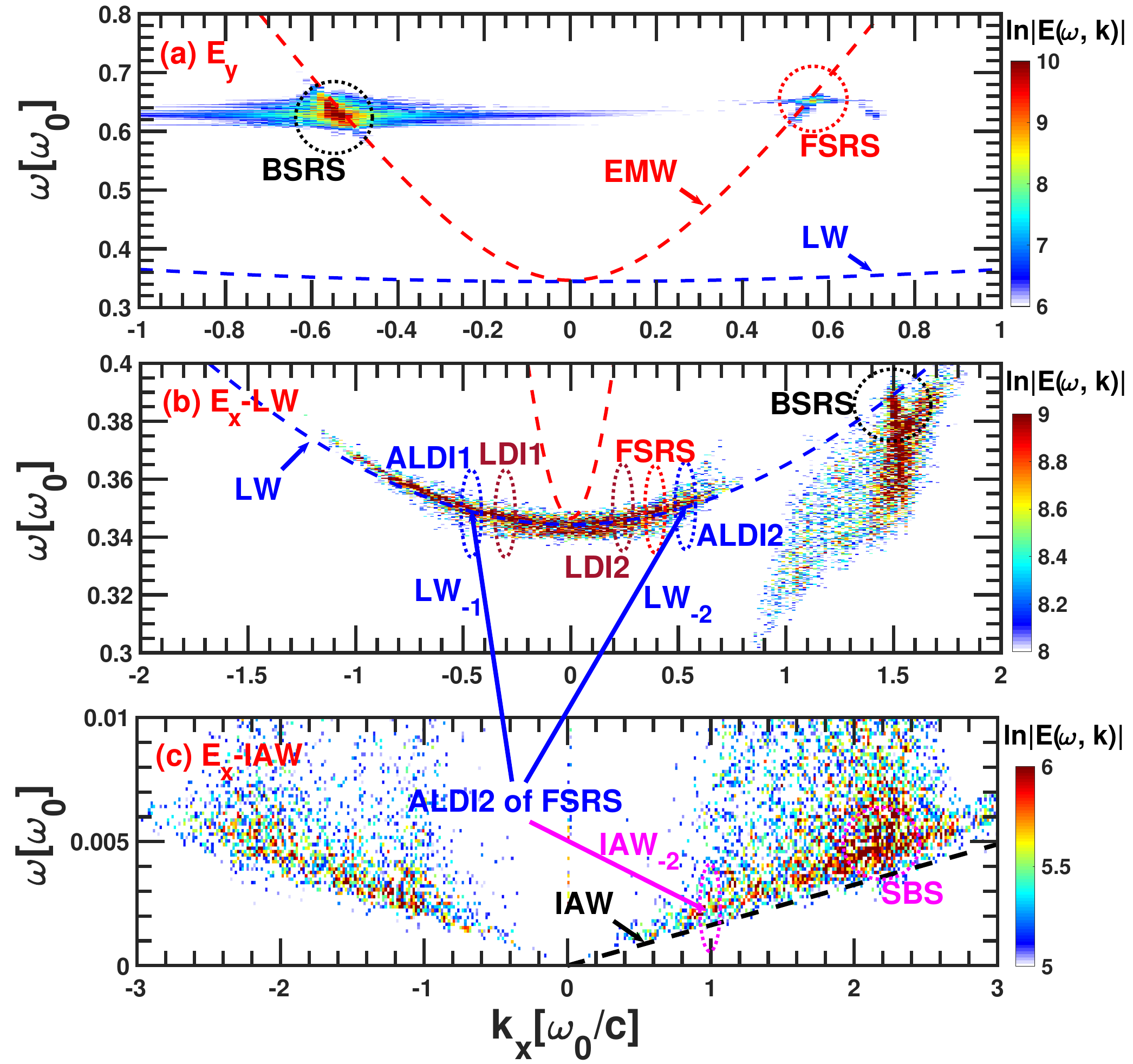}
		\caption{\label{Fig:w_k_ne=0.12}(Color online) The dispersion relation of (a) transverse electric field $E_y$ and (b), (c) longitudinal electric field $E_x$. Where (a)-(c) are during the time range $t\in[5000, 50000]\omega_0^{-1}$ and the space range $x\in[100, 400]c/\omega_0$. The parameters are $n_e=0.12n_c, T_e=\SI{2.5}{keV}$, $ I_0=1\times10^{16}\, \SI{}{W/cm^2}$ in H plasmas.}
	\end{figure}
	
	\subsubsection{Electron acceleration by anti-Stokes Langmuir decay instability cascade in Region II}
	When the electron density decreases from $n_e=0.2n_c$ (Fig. \ref{Fig:PhasePicture}), $n_e=0.15n_c$ (Fig. \ref{Fig:PhasePicture_ne=0.15}), $n_e=0.13n_c$ (Fig. \ref{Fig:PhasePicture_ne=0.13}) to $n_e=0.12n_c$ (Fig. \ref{Fig:PhasePicture_ne=0.12}) in descending order, the maximum momentum, and therefore the kinetic energy of the trapped electrons, decreases. As shown in Figs. \ref{Fig:PhasePicture_ne=0.15}(a) and \ref{Fig:PhasePicture_ne=0.15}(b), the electrons trapped and accelerated by the backward-SRS induced Langmuir wave are further trapped by the forward-SRS induced Langmuir wave, although the forward-SRS induced vortices shaping the electron phase spaces when $n_e=0.15n_c$ are less obvious than those when $n_e=0.2n_c$. 
	
	The threshold electron density ($n_e = 0.138 n_c$) beyond which electrons can be accelerated via the backward-SRS and the forward-SRS process is estimated assuming the backward-SRS induced Langmuir wave reaches the wave breaking limit, as analytically evaluated in Refs. \cite{Coffey_1971POF,Mori_1990PS}. Those waves, however, can saturate at a lower amplitude due to Langmuir decay instability \cite{Feng_2018POP,Hinkel_2004POP}, sideband-type instabilities \cite{Brunner_2004PRL,Brunner_2014POP}, so that the model only gives an approximate lower limit of the threshold density for backward-SRS and forward-SRS coupling. 
	There are indeed a few electrons with momentum larger than the phase velocity of the forward-SRS induced Langmuir wave, which verifies that the two stage acceleration of backward-SRS and forward-SRS occurs when the electron density is  $n_e=0.15n_c$ in Region III as shown in Fig. \ref{Fig:PhasePicture_ne=0.15}(b). 
	
	While in the Region II as shown in Fig. \ref{Fig:Theoretical model}(a), two cases for $n_e=0.13n_c$ and $n_e=0.12n_c$ are chosen as examples. As shown in Figs. \ref{Fig:PhasePicture_ne=0.13}(a)-(c) and Figs. \ref{Fig:PhasePicture_ne=0.12}(a)-(c), none of electrons are trapped and accelerated by the forward-SRS induced Langmuir wave. The maximum momenta of electrons at any time do not exceed that at the phase velocity of the forward-SRS induced Langmuir wave. The electron momentum remains at a low level. However, there are a small number of  electrons with momenta larger than $p_m^B$ (the maximum electron momentum accelerated by the backward-SRS induced Langmuir wave). These energetic electrons with momenta larger than $p_m^B$ is due to generation of Langmuir waves with higher phase velocity than the backward-SRS induced Langmuir wave, which is from the second stage anti-Stokes Langmuir decay instability (ALDI2) of forward-SRS, which will be clarified in the next paragraph.

	The negative propagation electrons are generated from the Langmuir wave of the first or third anti-Stokes Langmuir decay instability (ALDI1 or ALDI3) of forward-SRS, but not from the Langmuir decay instability of backward-SRS. When $n_e=0.12$ or $0.13n_c$, the Langmuir decay instability of backward-SRS is difficult excite, because this is in the kinetic regime \cite{Kline_2005PRL,Kline_2006POP}. The green lines in Figs. \ref{Fig:PhasePicture_ne=0.13}(c) and \ref{Fig:PhasePicture_ne=0.12}(c) represent the momenta at the phase velocity of the Langmuir wave from the Langmuir decay instability of backward-SRS, one sees that nearly no negative propagation electrons are trapped around those momenta.
	
	To confirm that there exists a Langmuir wave that has a higher phase velocity than the backward-SRS induced Langmuir wave, the dispersion relation is plotted in Fig. \ref{Fig:w_k_ne=0.12}. When $n_e=0.12n_c$, the wave-number of the backward-SRS induced Langmuir wave is $k_L^B\lambda_{De}=0.29$, which is in the kinetic regime \cite{Montgomery_2001PRL}. Thus, the Langmuir decay instability of backward-SRS is not allowed. However, the wave-number of the forward-SRS induced Langmuir wave is $k_L^F=0.39\omega_0/c=0.078\lambda_{De}^{-1}$, which means it is in the fluid regime \cite{Kline_2005PRL,Kline_2006POP}. Thus, the Langmuir decay instability cascade of the forward-SRS occurs more easily and the anti-Stokes process of the Langmuir decay instability is generated in the cascade. As shown in Fig. \ref{Fig:w_k_ne=0.12}(b), there is no Langmuir decay instability of backward-SRS induced Langmuir wave, and the cascade of the forward-SRS induced Langmuir wave is very large, consistent with the theory. The wave-number of Langmuir waves generated by the n-stage Langmuir decay instability (denoted as LDIn) of forward-SRS induced Langmuir wave ($k_L^F$) is given by
	\begin{equation}
	k_{Ln}\simeq(-1)^n(|k_L^F|-n\cdot\Delta k),
	\end{equation}
	where $\Delta k=\frac{2}{3}\frac{1}{\lambda_{De}}\frac{c_s}{v_{te}}$, and the negative sign represents the propagation direction of the Langmuir wave.  When $n_e=0.12n_c$ and $T_e=\SI{2.5}{keV}$, one obtains $\Delta k=0.0156\lambda_{De}^{-1}=0.0771\omega_0/c$. Thus, $k_{L1}=-0.31\omega_0/c$ from LDI1 of forward-SRS, $k_{L2}=0.23\omega_0/c$ from LDI2 of forward-SRS. The n-stage anti-Stokes LDI (ALDIn) of forward-SRS induced Langmuir wave is given by
	\begin{equation}
	k_{L-n}\simeq(-1)^n(|k_L^F|+n\cdot\Delta k),
	\end{equation}
	thus, $k_{L-1}=-0.46\omega_0/c$ and $k_{L-2}=0.54\omega_0/c$ from the anti-Stokes Langmuir decay instability 1 and 2 (ALDI1 and ALDI2) of forward-SRS, respectively. When the wave-number of ALDI2 is $k_{L-2}=0.54\omega_0/c$, the corresponding phase velocity, electron momenta and kinetic energy of the electrons are $v_{\phi-2}=0.65c$, $p_{\phi-2}=0.86m_{e}c$ and  $E_{\phi-2}=\SI{161}{keV}$ at this particular phase velocity. The portion of energetic electrons with momenta larger than $p_m^B=0.60m_ec$ ($E_m^B=\SI{84}{keV}$) shown in Fig. \ref{Fig:PhasePicture_ne=0.12}(c). The clearer results of this novel mechanism will be further demonstrated in Sections \ref{subsection:Vlasov_long_scale} and \ref{section:EPOCH}. This is due to generation of the Langmuir wave (labelled as LW$_{-2}$) from the second stage anti-Stokes Langmuir decay instability (ALDI2) of forward-SRS.

	\begin{figure}[!tp]
		%	\centering
		\includegraphics[width=1\columnwidth]{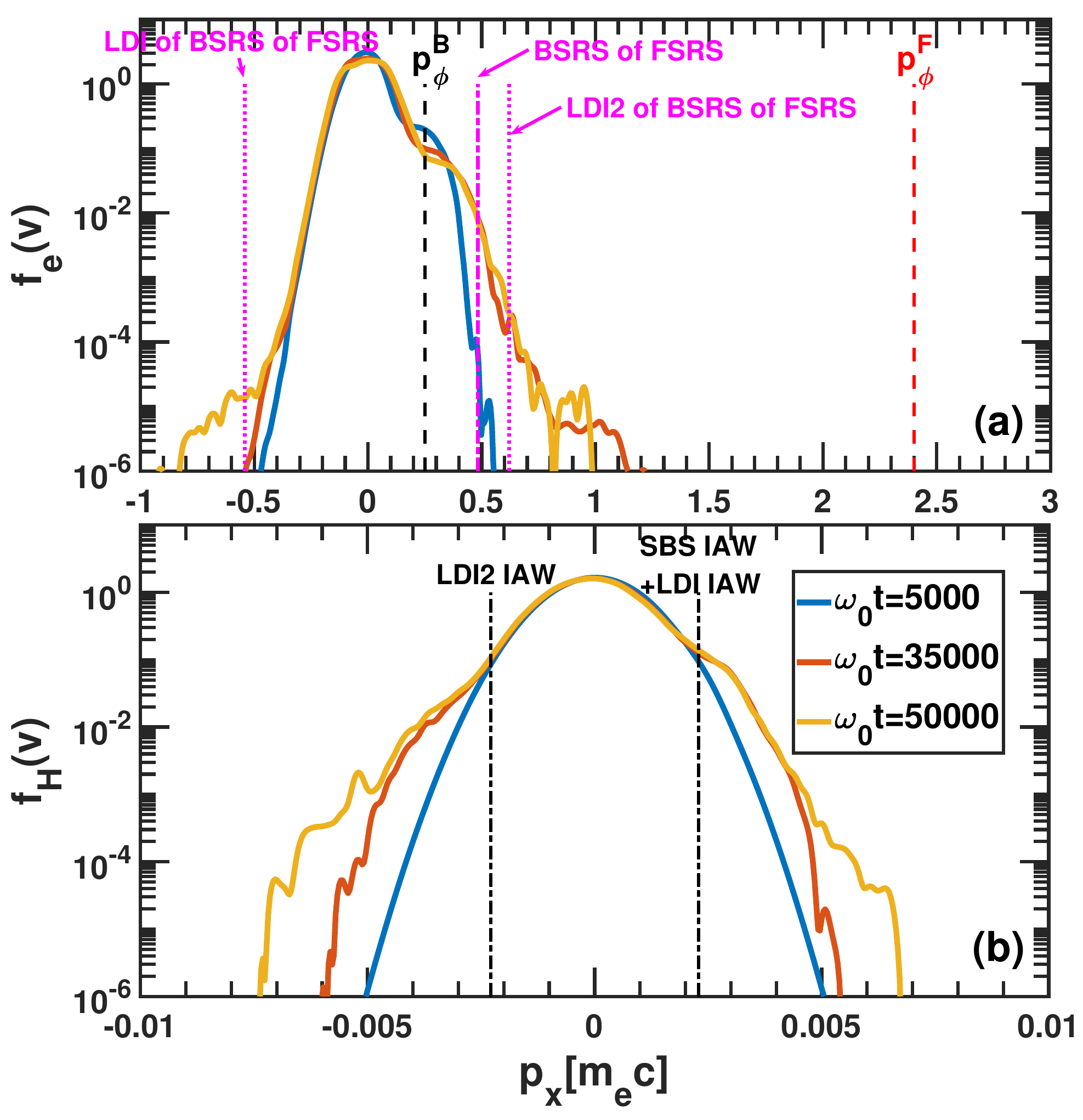}
		\caption{\label{Fig:ne=0.1nc}(Color online) The distributions of (a) electrons and (b) H ions averaged in total simulation space at different time. Where the simulation condition is $n_e=0.1n_c, T_e=\SI{2.5}{keV}$, $ I_0=5\times10^{16}\, \SI{}{W/cm^2}$ in H plasmas.
		}
	\end{figure}
	
	\subsubsection{Electron acceleration by rescattering and Langmuir decay instability cascade in Region I}
	As shown in Fig. \ref{Fig:Theoretical model}(a), $p_{min}^F>p_m^B$  when $n_e=0.1n_c, T_e=\SI{2.5}{keV}$ (Region I). Therefore, the electrons trapped by the backward-SRS induced Langmuir wave cannot be trapped and accelerated by the forward-SRS induced Langmuir wave directly. To verify this conclusion, we have conducted Vlasov simulations for $n_e=0.1n_c, T_e=\SI{2.5}{keV}, I_0=5\times10^{16}\, \SI{}{W/cm^2}$ as shown in Fig. \ref{Fig:ne=0.1nc}. The backward-SRS gain $G_R=3.4$ when $I_0=1\times10^{16}\, \SI{}{W/cm^2}$ is very small, and as such both backward-SRS and forward-SRS experience low growth in Vlasov simulations when there is no seed electromagnetic wave. Thus a large intensity $I_0=5\times10^{16}\, \SI{}{W/cm^2}$ is used in this case. The electrons trapped by the backward-SRS induced Langmuir wave cannot be trapped by the forward-SRS induced Langmuir wave indeed even when $\omega_0t=50000$. When $n_e=0.1n_c$, backward-SRS of backward-SRS cannot occur because it is not possible to satisfy the three-wave matching conditions, and the Langmuir decay instability of backward-SRS is difficult to be excited. Backward-SRS of forward-SRS, Langmuir decay instability of backward-SRS of forward-SRS, and  a second stage Langmuir decay instability (LDI2) of backward-SRS of forward-SRS are presented, as shown in Fig. \ref{Fig:ne=0.1nc}(a). These rescattering and Langmuir decay instability processes further accelerate electrons that are trapped by the backward-SRS induced Langmuir wave. However, in this section, the plasma length is not very large, thus forward-SRS, backward-SRS of forward-SRS and the Langmuir decay instability cannot develop to a significant level. As a result, the forward-SRS induced Langmuir wave is not able to trap low-momenta electrons. On the other hand, if the plasma density length is large enough, the rescattering will accelerate the electrons to sufficiently high energy to be trapped by the forward-SRS induced Langmuir wave \cite{Winjum_2013PRL}.  Stimulated Brillouin scattering (SBS) can also exist in this parameter regime. The corresponding ion acoustic waves generated by stimulated Brillouin scattering, the Langmuir decay instability or the second stage Langmuir decay instability (LDI2) flatten the ion distribution at their phase velocity, as shown in Fig. \ref{Fig:ne=0.1nc}(b).
	
	\subsection{\label{subsection:Vlasov_long_scale}Laser-plasma interactions with large length-scales}
	In this section, we will explore the effect of the spatial scale on these different processes using the Vlasov simulation tool described previously. The spatial scale was set at [0, $L_x$] and discretized with $N_x=50000$ spatial grid points and spatial step $dx=0.1c/\omega_0$. The spatial length was $L_x=5000c/\omega_0$ with $2\times5\%L_x$ vacuum layers with $2\times5\%L_x$ collision layers on either sides of the plasma column. The pump intensity was $I_0=3\times10^{15}\, \SI{}{W/cm^2}$. The other parameters were the same as those described in Section \ref{subsection:Vlasov_short_scale}.
	
	\begin{figure}[!tp]
		\includegraphics[width=1\columnwidth]{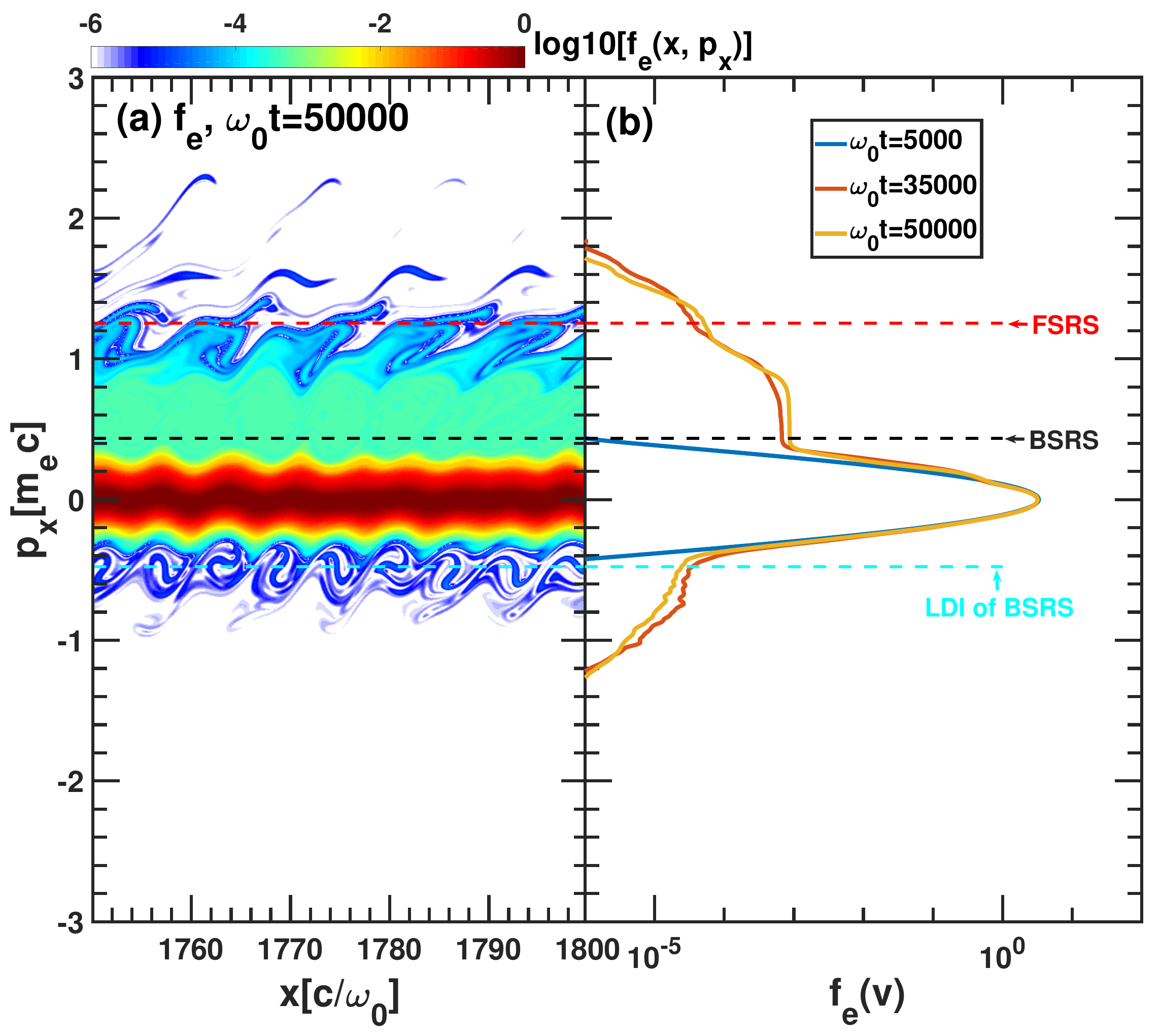}
		\caption{\label{Fig:px_x_ne=0.2}(Color online)
			(a) The phase space of electrons and (b) the corresponding electron distribution function averaged in space across $[1750, 1800]c/\omega_0$.   The parameters are $n_e=0.2n_c, T_e=\SI{2.5}{keV}, I_0=3\times10^{15}\, \SI{}{W/cm^2}$ in H plasmas with $L_x=5000c/\omega_0$.}
	\end{figure}
	
	\begin{figure}[!tp]
		%	\centering
		\includegraphics[width=1\columnwidth]{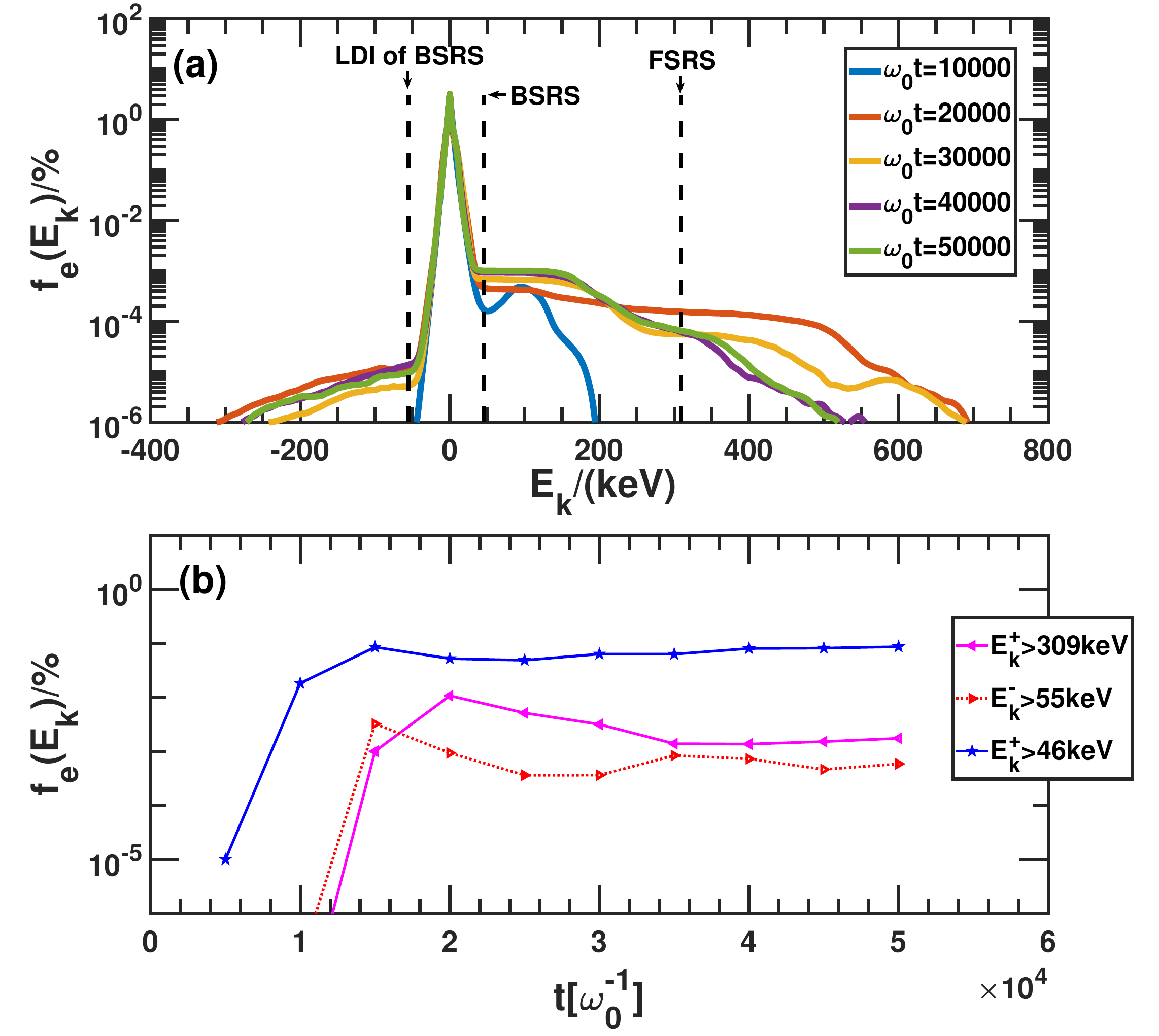}
		\caption{\label{Fig:f_Ek_ne=0.2}(Color online) (a) The distribution of electron kinetic energy at different times, spatially averaged over the range $[1000, 4000]c/\omega_0$. (b) The evolution in time of the ratios of the electrons in different energy ranges. The black dashed lines labelled in (a) represent the phase velocities of corresponding Langmuir waves. $E_k^{\pm}$ represents the kinetic energy of the electrons with positive velocity ($+$) and negative velocity ($-$). From theoretical considerations, $E_\phi^B=\SI{46}{keV}, E_\phi^F=\SI{309}{keV}$ and $E_\phi^L=\SI{55}{keV}$ from LDI of backward-SRS when $n_e=0.2n_c, T_e=\SI{2.5}{keV},  I_0=3\times10^{15}\, \SI{}{W/cm^2}, L_x=5000c/\omega_0$.
		}
	\end{figure}

	As shown in Fig. \ref{Fig:px_x_ne=0.2}, when $n_e=0.2n_c, L_x=5000c/\omega_0$ (Region III), the two stage acceleration by backward-SRS and forward-SRS occurs. This is similar to the results obtained for when the parameters were $n_e=0.2n_c, L_x=500c/\omega_0$, although the pump light intensity is $I_0=3\times10^{15}\, \SI{}{W/cm^2}$, somewhat lower than that in the short plasma simulations shown in Section \ref{subsection:Vlasov_short_scale}. Fig. \ref{Fig:f_Ek_ne=0.2}(a) shows that when $\omega_0t=1\times10^4$, only the backward-SRS induced Langmuir wave traps and accelerates electrons. After $\omega_0t\simeq2\times10^4$, the forward-SRS induced Langmuir wave traps and accelerates electrons to a maximum energy of approximately $\SI{700}{keV}$, which is slightly lower than the maximum limit energy $E_{max}^F=\SI{797}{keV}$ from analytic theory. At the same time, the  Langmuir wave generated from the Langmuir decay instability of backward-SRS traps electrons and produces negative propagation energetic electrons after $\omega_0t\simeq2\times10^4$. 
	
	The ratios of electrons with energy above the energy corresponding to phase velocities of different Langmuir waves from backward-SRS (\SI{46}{keV}), Langmuir decay instability of backward-SRS (\SI{55}{keV}) and forward-SRS (\SI{309}{keV}) are shown in Fig. \ref{Fig:f_Ek_ne=0.2}(b). The saturation level of energetic electrons with $E_k>\SI{46}{keV}$ accelerated by the backward-SRS induced Langmuir wave is $\sim0.1\%$, while the level of  energetic electrons with $E_k>\SI{309}{keV}$ accelerated by the forward-SRS induced Langmuir wave is $\sim0.01\%$. These results illustrate that the two stage acceleration of electrons by backward-SRS and forward-SRS occurs when $n_e=0.2n_c$ in both long ($L_x=5000c/\omega_0$) and short ($L_x=500c/\omega_0$) scale plasma.

\begin{figure}[!tp]
		\includegraphics[width=1\columnwidth]{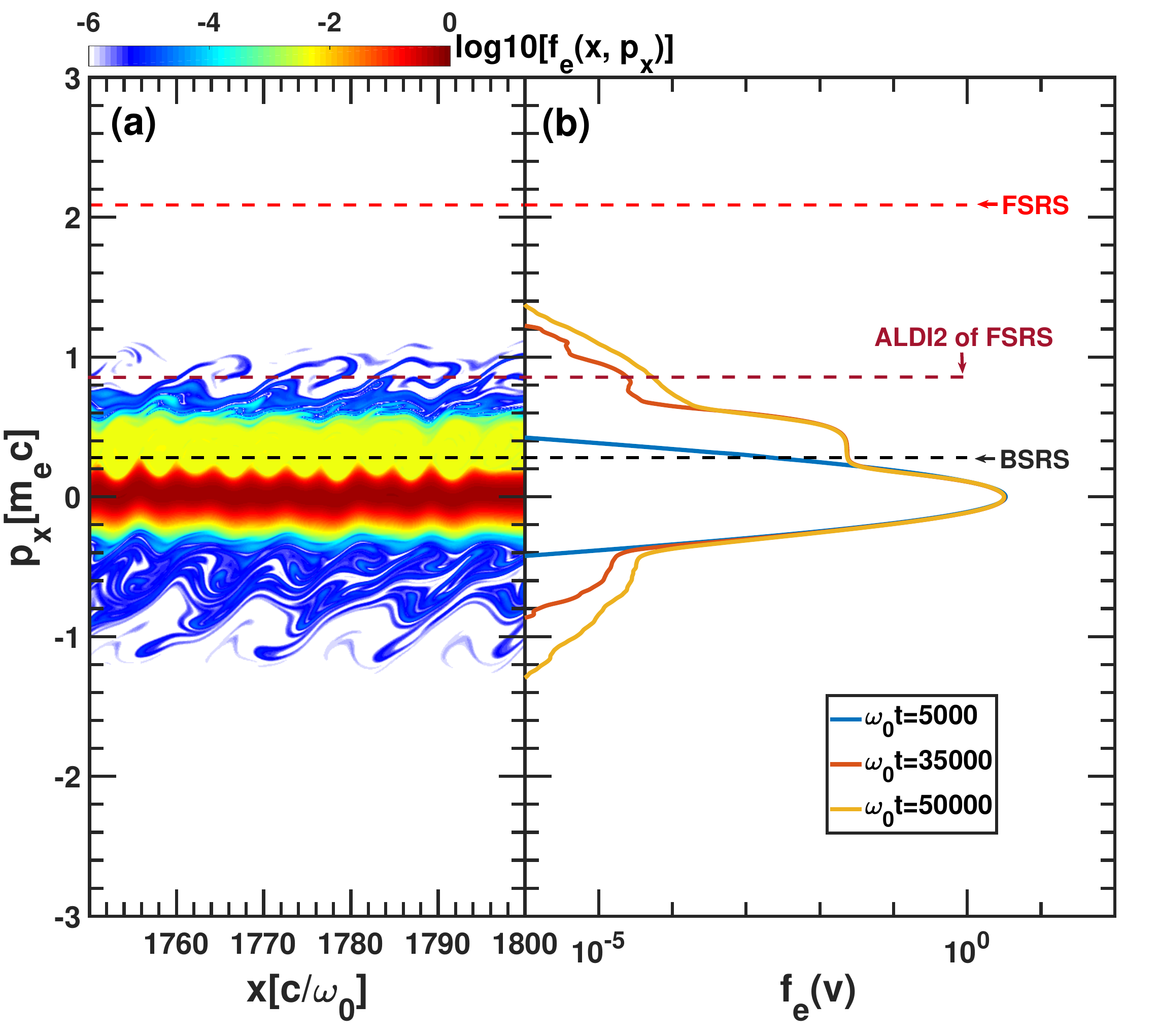}
		\caption{\label{Fig:px_x_ne=0.12}(Color online)
			(a) The phase space of electrons and (b) the corresponding electron distribution function averaged in space across $[1750, 1800]c/\omega_0$.   The parameters are $n_e=0.12n_c, T_e=\SI{2.5}{keV}, I_0=3\times10^{15}\, \SI{}{W/cm^2}$ in H plasmas with $L_x=5000c/\omega_0$.}
	\end{figure}

	Let us examine what happens when the electron density $n_e=0.12n_c$, i.e. in Region II (shown in Fig. \ref{Fig:Theoretical model}(a)). Fig. \ref{Fig:px_x_ne=0.12} shows that the electrons are trapped and accelerated by the backward-SRS induced Langmuir wave, but cannot be trapped by the forward-SRS induced Langmuir wave due to a large gap between the phase velocities of the forward-SRS and backward-SRS induced Langmuir waves. 
	
	The maximum momentum of electrons trapped by the backward-SRS induced Langmuir wave reaches only $p_m^B=0.60m_ec$ from the wave-breaking limit. However, there is a tail of energetic electrons with momenta larger than $p_m^B$. These electrons are trapped and accelerated by a  Langmuir wave with a higher phase velocity arising from the second stage anti-Stokes Langmuir decay instability ALDI2 of forward-SRS, as demonstrated above in Fig. \ref{Fig:w_k_ne=0.12}. 
	
	The momentum at phase velocity of the Langmuir wave is $p_{\phi-2}=0.86m_ec$, which is indeed at the energetic electron tail. Although electrons are accelerated by backward-SRS and ALDI2 of forward-SRS to momenta of nearly $p_m\sim1.4m_ec$, this is also much lower than the momenta at the phase velocity of the forward-SRS $p_\phi^F=2.1m_ec$. Since there exists a number of saturation mechanisms of both backward-SRS and forward-SRS, such as sideband instability of backward-SRS \cite{Brunner_2004PRL,Brunner_2014POP} and Langmuir decay instability of forward-SRS \cite{Feng_2018POP,Hinkel_2004POP}, there is a maximum electric field that cannot approach the wave-breaking limit. Therefore, the electrons accelerated by ALDI2 of forward-SRS cannot be trapped and accelerated by forward-SRS as shown in Fig. \ref{Fig:px_x_ne=0.12}. 
	
\begin{figure}[!tp]
		%	\centering
		\includegraphics[width=1\columnwidth]{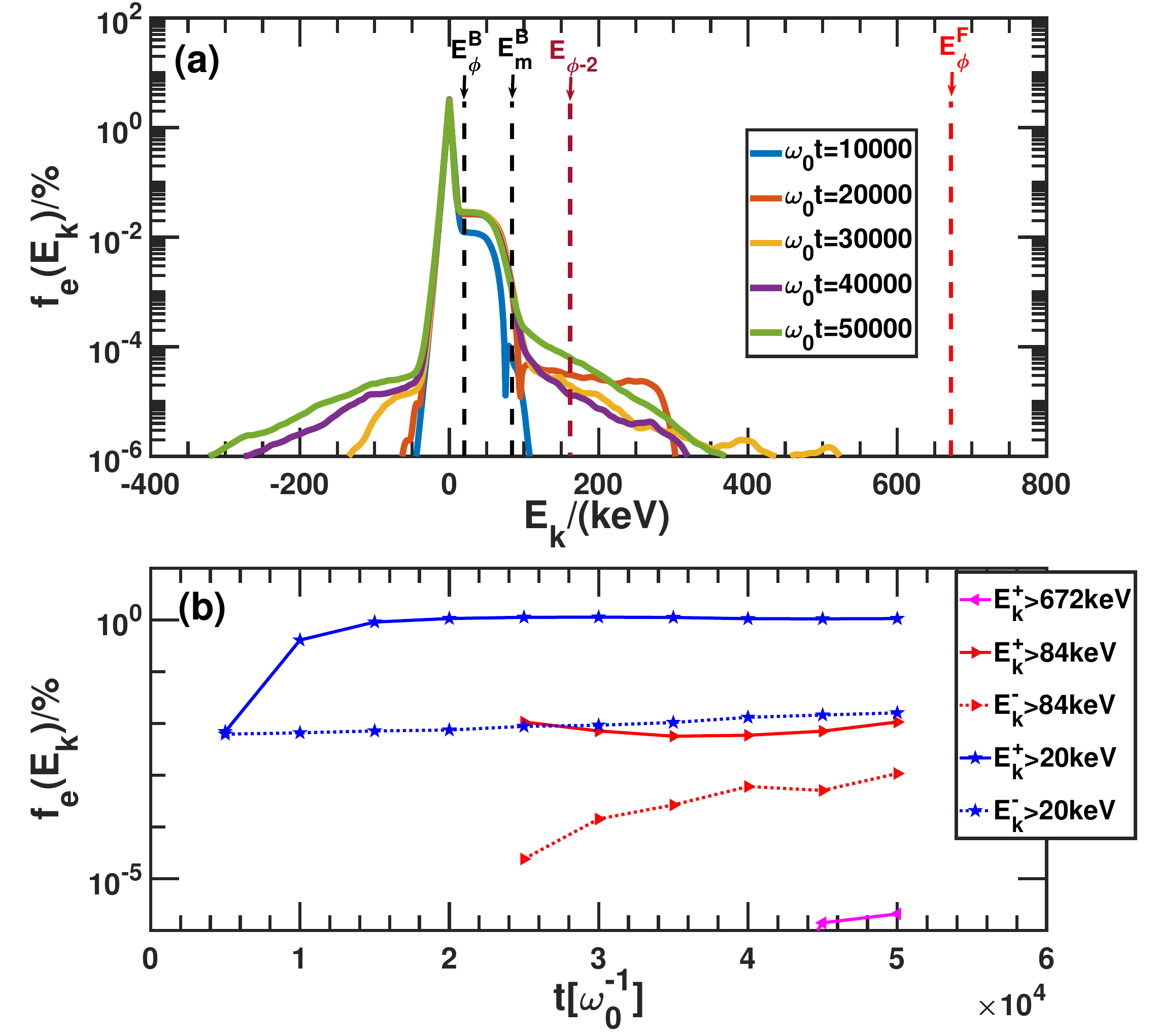}
		\caption{\label{Fig:f_Ek_ne=0.12}(Color online) (a) The distribution of electron kinetic energy at different times. (b) The evolution in time of the ratios of the electrons in different energy ranges. The black dashed lines labelled in (a) represent the phase velocities of corresponding  Langmuir waves. $E_k^{\pm}$ represents the kinetic energy of electrons with positive velocity ($+$) and negative velocity ($-$). From theoretical calculation, $E_\phi^B=\SI{20}{keV}, E_m^B=\SI{84}{keV}, E_{\phi-2}=\SI{161}{keV}, E_\phi^F=\SI{672}{keV}$ when $n_e=0.12n_c, T_e=\SI{2.5}{keV},  I_0=3\times10^{15}\, \SI{}{W/cm^2}, L_x=5000c/\omega_0$.
		}
	\end{figure}

	From the energy spectrum shown in Fig. \ref{Fig:f_Ek_ne=0.12}(a), the electrons are solely trapped and accelerated by the backward-SRS induced Langmuir wave before $\omega_0t\sim1\times10^4$. When $\omega_0t=2\times10^4$, more energetic electrons are generated around $E_{\phi-2}=\SI{161}{keV}$. This is the  Langmuir wave from ALDI2 of forward-SRS. The maximum energy reaches nearly $\SI{400}{keV}$, but the electrons can not yet be trapped and accelerated by the forward-SRS induced Langmuir wave ($E_\phi^F=\SI{672}{keV}$). 
	
	The evolution of electron ratios with different energy windows is shown in Fig. \ref{Fig:f_Ek_ne=0.12}(b). After $\omega_0t\sim1\times10^{4}$, the electrons with energy above $E_\phi^B=\SI{20}{keV}$ nearly saturate, which illustrates that backward-SRS saturates after $\omega_0t\sim1\times10^4$. Also, the electrons with energy above $E_m^B=\SI{84}{keV}$ are generated from the Langmuir wave of ALDI2 of forward-SRS, the ratio of which is about $0.01\%$. Finally, the electrons with energy above $E_\phi^F=\SI{672}{keV}$ result in a ratio nearly equal to zero and can therefore be neglected.
	
	\begin{figure}[!tp]
		\includegraphics[width=1\columnwidth]{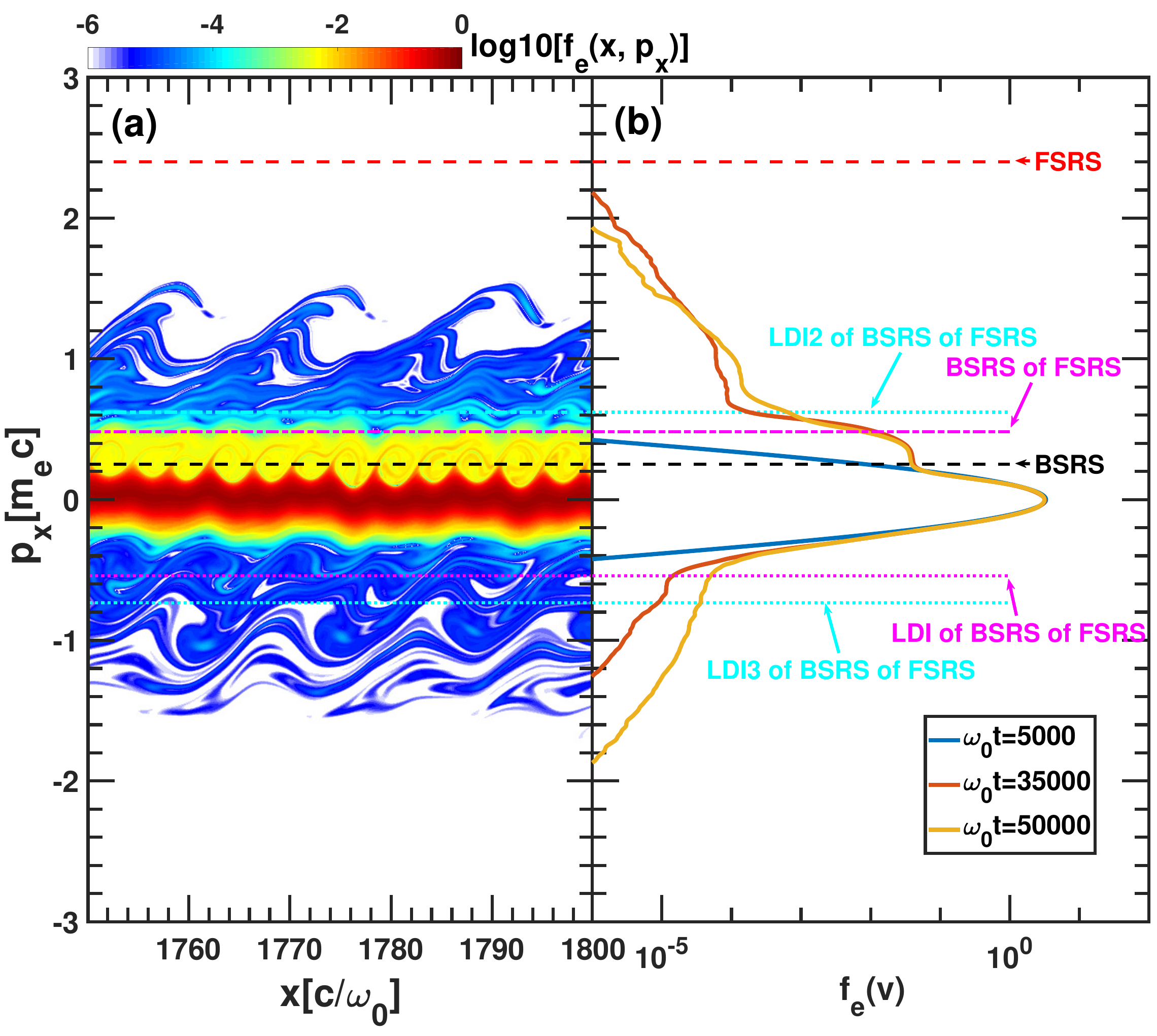}
		\caption{\label{Fig:px_x_ne=0.1}(Color online)
			(a) The phase space of electrons and (b) the corresponding electron distribution function averaged in space across $[1750, 1800]c/\omega_0$.   The parameters are $n_e=0.1n_c, T_e=\SI{2.5}{keV}, I_0=3\times10^{15}\, \SI{}{W/cm^2}$ in H plasmas with $L_x=5000c/\omega_0$.}
	\end{figure}
	
	\begin{figure}[!tp]
		%	\centering
		\includegraphics[width=1\columnwidth]{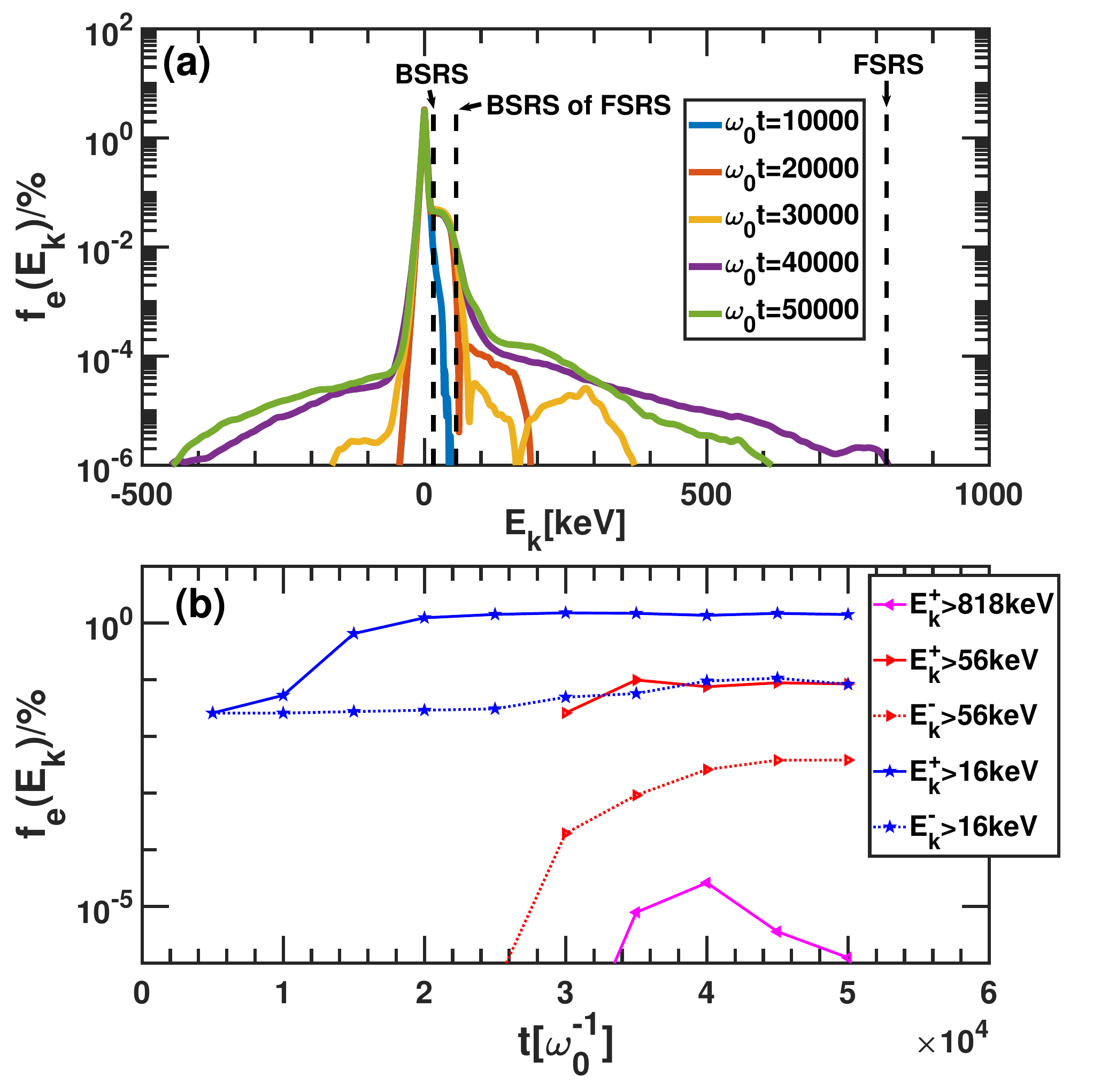}
		\caption{\label{Fig:f_Ek_ne=0.1}(Color online) (a) The distribution of electron kinetic energy at different times. (b) The evolution in time of the ratios of the electrons in different energy ranges. The black dashed lines labelled in (a) represent the kinetic energy at the phase velocities of corresponding Langmuir waves. $E_k^{\pm}$ represents the kinetic energy of electrons with positive velocity ($+$) and negative velocity ($-$). From theoretical calculation, $E_\phi^B=\SI{16}{keV}, E_\phi^F=\SI{818}{keV}$ and $E_\phi^r=\SI{56}{keV}$ from the rescattering backward-SRS of forward-SRS when $n_e=0.1n_c, T_e=\SI{2.5}{keV}, I_0=3\times10^{15}\, \SI{}{W/cm^2}$ and $L_x=5000c/\omega_0$.
		}
	\end{figure}

	Let us now turn our attention to the situation when the electron density is  $n_e=0.1n_c$ in order to illustrate what happens in Region I (shown in Fig. \ref{Fig:Theoretical model}(a)).  It is clear that the laser pulse interaction with large scale plasma produces rescatterings and corresponding Langmuir decay instabilities, as shown in Fig. \ref{Fig:px_x_ne=0.1}.
	
	Here the phase velocity of the forward-SRS induced Langmuir wave is now much higher than that of the backward-SRS induced Langmuir wave. This means that the electrons accelerated by the backward-SRS induced Langmuir wave cannot be trapped and accelerated by the forward-SRS induced Langmuir wave directly. Rather, under these particular conditions, only backward-SRS of forward-SRS occurs. Since the wave number of the backward-SRS induced Langmuir wave is $k_L^B\lambda_{De}=0.33$, this means that the process is in the kinetic regime. Thus the Langmuir decay instability of backward-SRS is marginal for growth. 
	
	On the other hand,  the wave number of backward-SRS of forward-SRS induced Langmuir wave is $k_L^r\lambda_{De}=0.17$. This is in the fluid regime. Thus the Langmuir decay instability cascade associated with this  rescattering does occur more easily. At the same time, the Langmuir decay instability cascade of forward-SRS induced Langmuir wave ($k_L^F\lambda_{De}=0.076$) also grows more easily. 
	
	The backward-SRS of forward-SRS and Langmuir decay instability cascade of backward-SRS of forward-SRS produce  Langmuir waves with higher phase velocities than the backward-SRS induced Langmuir waves. Therefore, the cascade process provides the mechanism for electrons to be accelerated from low to high energy. 
	
	The maximum momenta that electrons can acquire is nearly $p_x\sim2.2m_ec$, as illustrated in Fig. \ref{Fig:px_x_ne=0.1}(b), and the maximum energy is  $E_k\sim\SI{820}{keV}$, as shown in Fig. \ref{Fig:f_Ek_ne=0.1}(a). This is close to $E_\phi^F=\SI{818}{keV}$.This provides conclusive evidence that when the electron density is $n_e\lesssim0.108n_c$ (Region I), the cascade acceleration mechanism by backward-SRS, rescatterings, Langmuir decay instability of rescatterings and forward-SRS occur \cite{Winjum_2013PRL}. These instabilities also accelerate electrons to high energy.
	
	\section{\label{section:EPOCH}1D and 3D Particle-in-Cell simulations}

	The one-dimensional (1D)  and three-dimensional (3D) Particle-in-Cell (PIC) code EPOCH \cite{Arber_2015PPCF,Feng_2020NF} was used to investigate the hot electron generation mechanisms from the different instabilities to confirm the veracity of the results in Section \ref{section:II} when extrapolated to higher dimensions (this was not possible with the Vlasov code due to the available computational resources).
	
	The electron/ion and laser pulse conditions were the same as that in section \ref{subsection:Vlasov_long_scale}, i.e. the electron temperature was $T_e=2.5keV$ and electron density is $n_e=0.1, 0.12, 0.2n_c$, where $n_c$ was the critical density of the incident pump light. The ion temperature was $T_i=T_e/3$, the same as that in Vlasov simulations. The pump laser pulse intensity was $I_0=3\times10^{15}\, \SI{}{W/cm^2}$ with a plane wave envelope and linear polarization in $y$ direction for both 1D and 3D simulations. 
	
	In the 1D simulations, the spatial domain along x direction was set at [0, $800\lambda_0$] discretized with $N_x=2\times10^4$ spatial grid points and spatial step $dx=0.04\lambda_0$. The particles occupied the full simulation space of the box with no vacuum layer on either side of the plasma slab. The total simulation time was $t_{end}=8000T_0=\SI{9.4}{ps}$, where $T_0=\SI{1.17}{fs}$ was the period of $3\omega$ pump light (\SI{351}{nm}). 
	
	In 3D simulations, the simulation parameters were the same as those in the 1D simulations, but the spatial domains in y and z directions were [-$5\lambda_0$, $5\lambda_0$] and [-$5\lambda_0$, $5\lambda_0$] discretized with $N_y=10$ and $N_z=10$ spatial grid points. Open boundary conditions for the laser and thermal particles were used in the $x$-direction for both 1D and 3D simulations, and periodic boundaries were used in $y$ and $z$ directions for the 3D simulations. There were 1000 electrons and 1000 ions per cell in the 1D simulations and 10 electrons/ions per cell in the 3D simulations. 
	
\begin{figure}[!tp]
		\includegraphics[width=1\columnwidth]{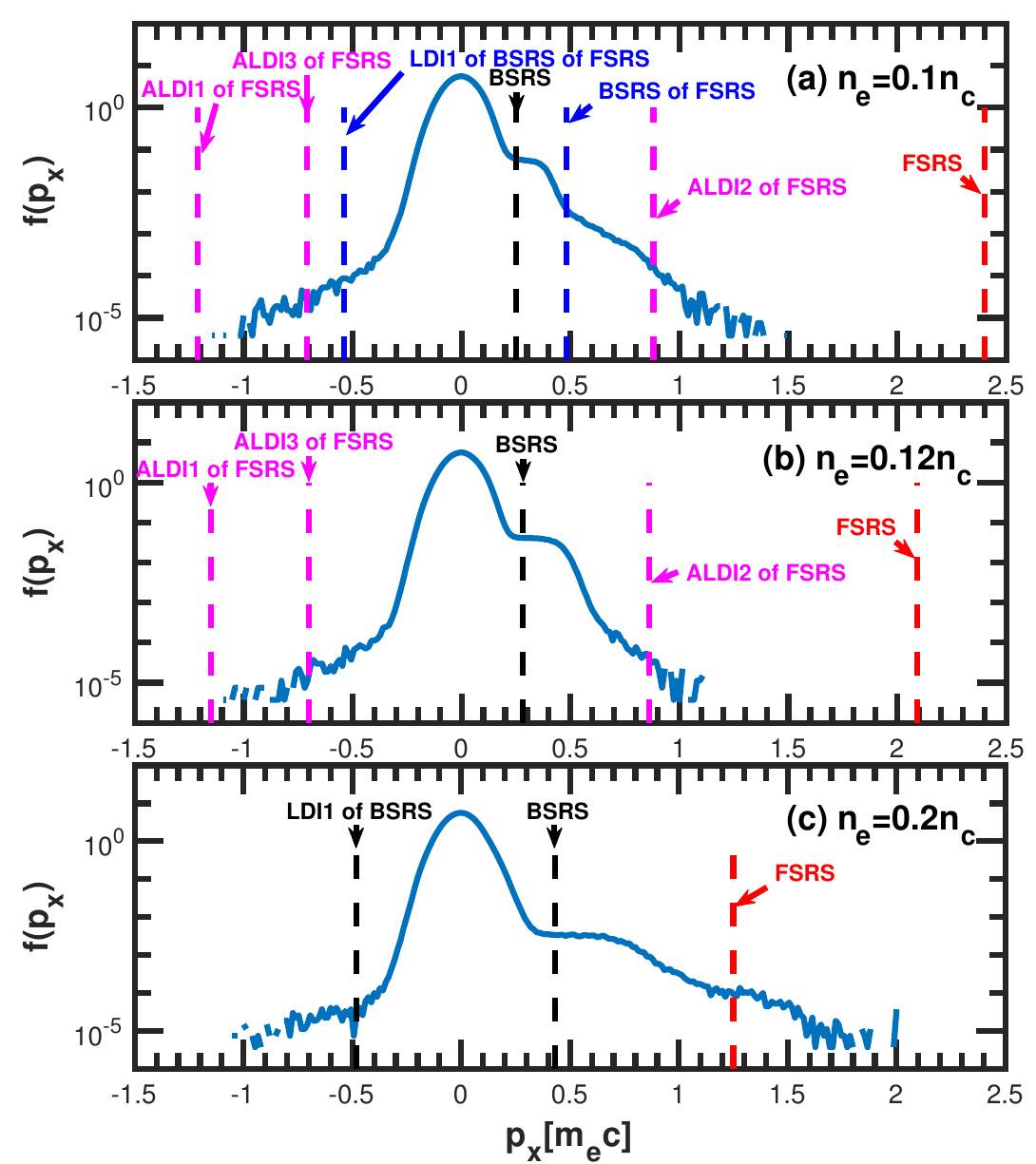}
		\caption{\label{Fig:1D_EPOCH}(Color online) The electron distribution functions from 1D EPOCH simulations when (a) $n_e=0.1n_c$, (b) $n_e=0.12n_c$, and (c) $n_e=0.2n_c$ at the time of $t=2000T_0$, which are spatially averaged across the whole length of the simulation domain along the x-direction $[0, 800 \lambda_0]$. The other parameters are $T_e=\SI{2.5}{keV}, I_0=3\times10^{15}\, \SI{}{W/cm^2}$ in H plasmas with $L_x=800\lambda_0$.
		}
	\end{figure}

	Figure \ref{Fig:1D_EPOCH} shows the 1D simulations results for $n_e=0.1, 0.12$ and $0.2n_c$. The parameters were the same as those described in Section \ref{subsection:Vlasov_long_scale}. When $n_e=0.1n_c$ (Region I) shown in Fig. \ref{Fig:1D_EPOCH}(a), the electrons are firstly trapped by the backward-SRS induced Langmuir wave, then mainly accelerated by the Langmuir waves from backward-SRS of forward-SRS and the corresponding Langmuir decay instability cascade of this rescattering. At the same time, the Langmuir waves induced by anti-Stokes Langmuir decay instability cascade of forward-SRS are also shown in Fig. \ref{Fig:1D_EPOCH}(a), which promote the acceleration of electrons.
	
	When $n_e=0.12n_c$ (Region II), shown in Fig. \ref{Fig:1D_EPOCH}(b), the electrons are trapped and accelerated by the backward-SRS, anti-Stokes Langmuir decay instability cascade of forward-SRS induced Langmuir waves. The phase velocities of Langmuir waves induced by these instabilities labelled in Fig. \ref{Fig:1D_EPOCH} are shown in Table \ref{table1}. 
	
	When $n_e=0.2n_c$ (Region III), shown in Fig. \ref{Fig:1D_EPOCH}(c), the electrons are trapped and accelerated by backward-SRS and forward-SRS directly, and the negative propagating electrons are accelerated by the Langmuir decay instability of backward-SRS. This is because no rescatterings exist for these conditions. Also, both backward-SRS and forward-SRS are much stronger than corresponding Langmuir decay instability. Consequently, the principal acceleration mechanisms are dual-stage electron acceleration by backward-SRS and forward-SRS. The results  from the 1D PIC simulations are close to those obtained by 1D Vlasov simulations, as shown in Section \ref{subsection:Vlasov_long_scale}, which illustrate that the results are robust between the simulation methods.
	
	\begin{figure}[!tp]
		\includegraphics[width=1\columnwidth]{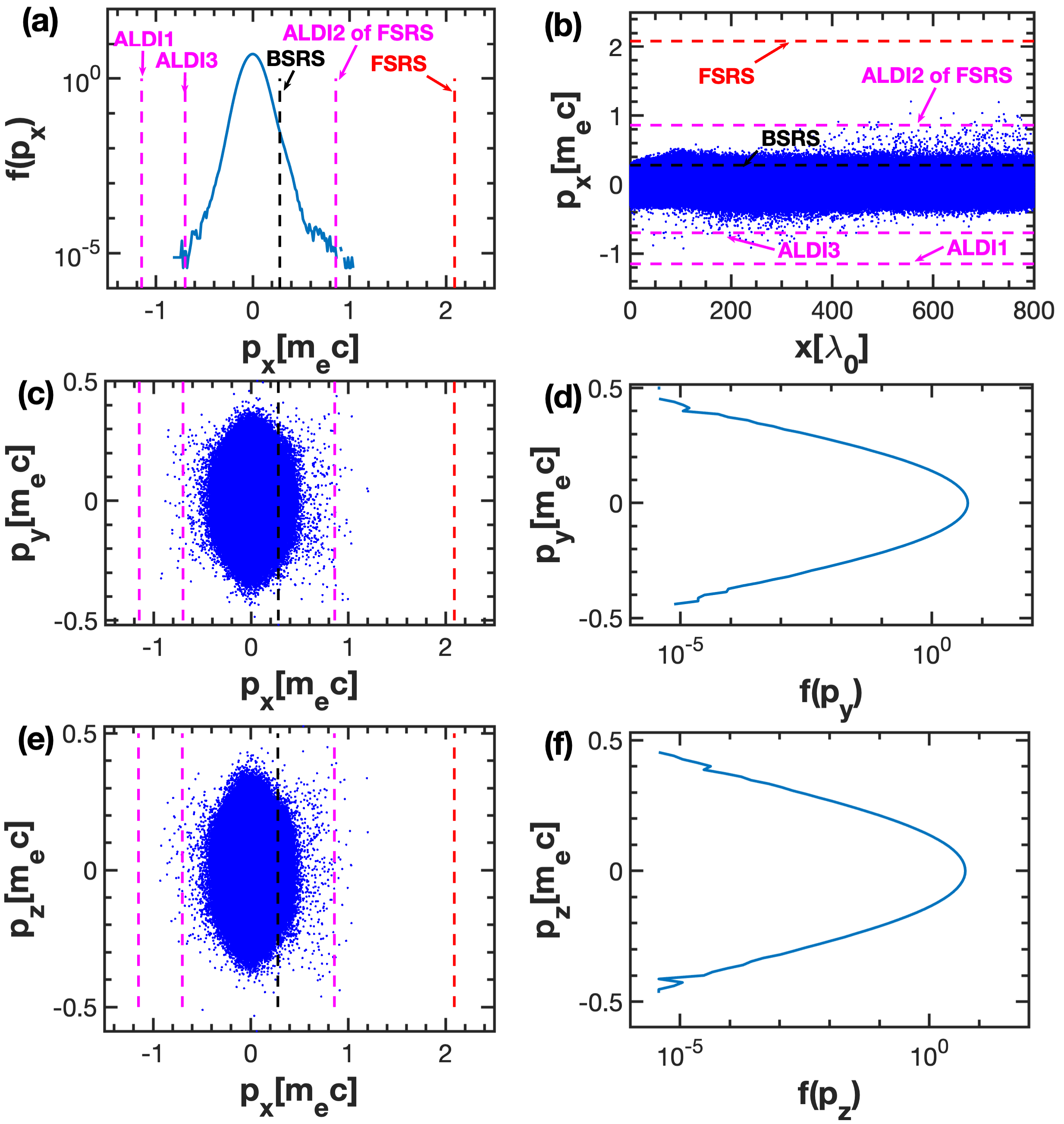}
		\caption{\label{Fig:3D_EPOCH}(Color online) The electron distribution of (a) $p_x$, (d) $p_y$ and (f) $p_z$ from 3D EPOCH simulations, which are averaged across the whole length of the simulation's domain along the x-direction $[0, 800 \lambda_0]$. (b) The phase space ($x-px$) of electrons. The electron momentum distribution across (c) $x$ and $y$, (e) $x$ and $z$. The time is $t=5600T_0$ and the parameters are $n_e=0.12n_c, T_e=\SI{2.5}{keV}, I_0=3\times10^{15}\, \SI{}{W/cm^2}$ for the H plasmas considered in the simulations.
		}
	\end{figure}
	
	To further understand the electron acceleration processes in 3D, multi-dimensional PIC simulations were also performed. Figure \ref{Fig:3D_EPOCH} shows the electron distributions of momenta $p_x$, $p_y$ and $p_z$ when $n_e=0.12n_c$ (Region II). Figure \ref{Fig:3D_EPOCH}(a) demonstrates that the electrons are trapped by the backward-SRS induced Langmuir wave and there are higher momenta tails in the electron distribution.  The positive momenta correspond to the phase velocity of the Langmuir wave from the second stage anti-Stokes Langmuir decay instability ALDI2 of forward-SRS, while the negative momenta are associated with the third stage anti-Stokes Langmuir decay instability ALDI3 of forward-SRS. This illustrates that the anti-Stokes Langmuir decay instability cascade of forward-SRS is the precise mechanism to accelerate electrons to higher momenta when $n_e=0.12n_c$. 
	
	The snapshots of the electron distribution (Fig. \ref{Fig:3D_EPOCH}(a)) and the electron phase space ($x-px$) (Fig. \ref{Fig:3D_EPOCH}(b)) are nearly identical to those generated by both the 1D EPOCH simulations (Fig. \ref{Fig:1D_EPOCH} (b)) and the 1D Vlasov simulations (Fig. \ref{Fig:px_x_ne=0.12}). This confirms that the mechanism of electron acceleration by anti-Stokes Langmuir decay instability cascade of forward-SRS is robust. 
	
	Figures \ref{Fig:3D_EPOCH}(c)-\ref{Fig:3D_EPOCH}(f) show the momenta distributions of electrons in both the $y$ and $z$ directions. The electrons have a near-Maxwellian distribution and are not accelerated in the transverse directions ($y$ and $z$ directions). This illustrates that the main acceleration mechanism is in the laser propagating direction ($x$ direction), and the novel mechanism of electron acceleration by anti-Stokes Langmuir decay instability cascade of forward-SRS is the principal acceleration mechanism to generate hot electrons with energy above $\SI{100}{keV}$ in Region II. 
	
	\section{\label{section:Discussions}Discussion}
	The maximum momenta of electrons from three typical cases when $n_e=0.1n_c, 0.12n_c, 0.2n_c$ in Region I, II, and III are shown in Fig. \ref{Fig:Theoretical model}(a). As discussed above, when $n_e=0.1n_c$ in Region I, the electrons are accelerated by backward-SRS, backward-SRS of forward-SRS and the Langmuir decay instability of rescattering to a maximum energy of nearly $\SI{820}{keV}$. Then, when $n_e=0.12n_c$ in Region II, the electrons are accelerated by backward-SRS and the second stage anti-Stokes Langmuir decay instability (ALDI2) of forward-SRS to a maximum energy of nearly $\SI{400}{keV}$. Finally, when $n_e=0.2n_c$ in Region III, the electrons are accelerated by backward-SRS and forward-SRS directly to a maximum energy of nearly $\SI{700}{keV}$.
	
	The ratios of electrons with energy above $E_m^B$ in Region I and Region III are obviously higher than that in Region II (only $\sim0.01\%$ in Region II). 
	These results illustrate that the optimal density region when $T_e=\SI{2.5}{keV}$ is $0.108n_c\lesssim n_e\lesssim 0.138n_c$ in Region II, in which the suprathermal electrons are reduced to a lower energy and lower ratio for indirect drive ICF experiments.
	
\begin{figure}[!tp]
		\includegraphics[width=1\columnwidth]{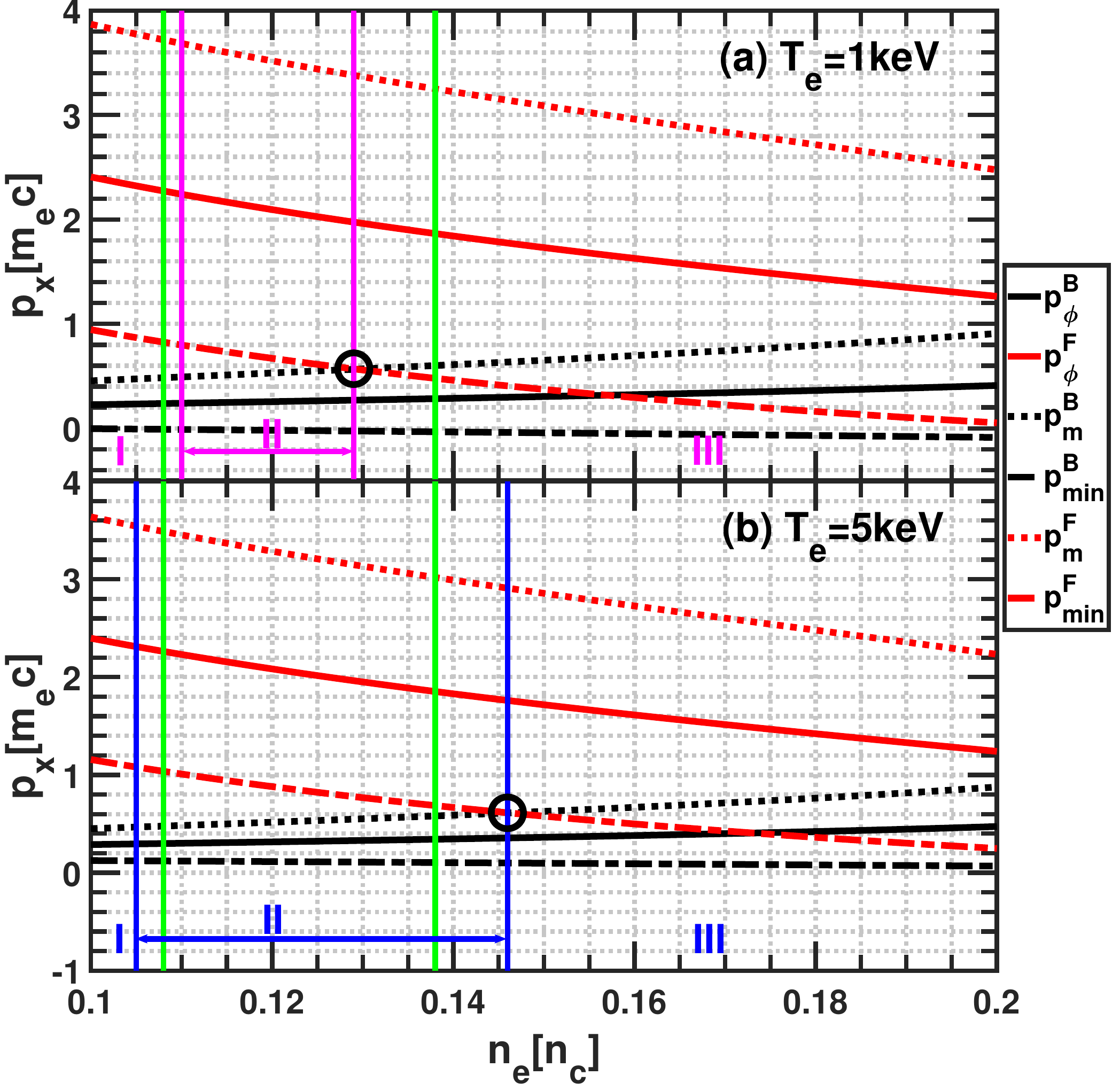}
		\caption{\label{Fig:Theoretical model_Te=1_Te=5}(Color online) The variation of $p_x$ with the electron density $n_e$ when (a) $T_e=\SI{1}{keV}$ and (b) $T_e=\SI{5}{keV}$. The green lines represent the upper and lower density boundaries of Region II when $T_e=\SI{2.5}{keV}$.
		}
	\end{figure}

	When the electron temperature varies, the upper and lower electron density boundaries of Region II also vary, as shown in Fig. \ref{Fig:Theoretical model_Te=1_Te=5}. When $T_e=\SI{1}{keV}$, the electron density range in Region II is $0.110n_c\lesssim n_e\lesssim 0.129n_c$ shown in Fig. \ref{Fig:Theoretical model_Te=1_Te=5}(a). As the temperature rises to $T_e=\SI{2.5}{keV}$, the electron density is bounded in Region II between $0.108n_c\lesssim n_e\lesssim 0.138n_c$, as shown in Fig. \ref{Fig:Theoretical model}(a), which is also marked in Figs. \ref{Fig:Theoretical model_Te=1_Te=5}(a) and \ref{Fig:Theoretical model_Te=1_Te=5}(b). 
	
	When the electron temperature rises still further to $T_e=\SI{5}{keV}$, the electron density in Region II is bounded by $0.105n_c\lesssim n_e\lesssim 0.146n_c$, as shown in Fig. \ref{Fig:Theoretical model_Te=1_Te=5}(b). That is to say, when $T_e$ varies from $\SI{1}{keV}$ to $\SI{5}{keV}$, which is the common electron temperature range in ICF experiments, the lower electron density boundary in Region II varies from $0.110n_c$ to $0.105n_c$ and the upper electron density boundary of Region II varies from $0.129n_c$ to $0.146n_c$. 
	
	This means that the boundaries of Region II change slightly when the electron temperature is in the range $[1, 5]$ $\SI{}{keV}$. Thus, our model for the electron density in Region II can be used to control the suprathermal electrons to be at a reduced level, since the suprathermal electrons are not sensitive to the electron temperature, as shown in Fig. \ref{Fig:Theoretical model}(b). Since those Langmuir waves from backward-SRS and forward-SRS, however, saturate at a lower amplitude due to secondary \cite{Feng_2018POP,Hinkel_2004POP}, sideband-type instabilities \cite{Brunner_2004PRL,Brunner_2014POP}, the model only gives an approximate lower limit of the threshold density for backward-SRS and forward-SRS coupling. That is to say that the model given in this paper predicts the safest region of electron density (Region II) to restrict the suprathermal electrons to be at a minimum level, where the two stage electron acceleration process by backward-SRS and forward-SRS does not occur.
	
	All of the calculations presented in this work assumed homogeneous plasma. For indirect-drive ICF, the hohlraum is filled with quasi-homogeneous low-density gas, so our work can be easily applied to NIF conditions by filling the hohraum with a gas at the appropriate density described in this work. In the case of inhomogeneous plasma, an effective damping from density gradient exists such that the forward-SRS will be suppressed compared to that in homogeneous plasma. Therefore, it is likely that the two stage acceleration of electrons by the backward-SRS and the forward-SRS will be much reduced in inhomogeneous plasma. We have conducted the simulations in inhomogeneous plasma with a density linearly increasing from $0.1n_c$ to $0.2n_c$, and found that the main mechanism of electron acceleration is the two-stage acceleration by backward-SRS and forward-SRS in high density region (such as $n_e=0.2n_c$). In this particular inhomogeneous case, the spectra of backward-SRS and forward-SRS are broad with a large bandwidth. The ratio of hot electrons above \SI{100}{keV} is less than that when $n_e=0.2n_c$ in homogeneous plasma. More detailed studies are left for future research.

	\section{\label{section:conclusions}Conclusion}
	In summary, a variety of different mechanisms in three distinct electron density regimes have been analyzed. A novel mechanism to accelerate electrons - anti-Stokes Langmuir decay instability cascade of forward-SRS - has been proposed to explain the generation of high energy electrons in Region II, which cannot be explained by traditional mechanisms. The wave-breaking maximum electric field has been used to predict the necessary condition for the two-stage electron acceleration by backward-SRS and forward-SRS and to distinguish the regions within which different electron acceleration mechanisms are allowed. We have shown that in Region I, with density $n_e\lesssim0.108n_c$, the rescattering and Langmuir decay instability of rescattering is an intermediate process which accelerates electrons trapped by the backward-SRS induced Langmuir wave. In Region II, when the density is in the range  $0.108n_c\lesssim n_e\lesssim 0.138n_c$, the ALDI2 (or second-stage anti-Stokes Langmuir decay instability cascade) of forward-SRS is the intermediate process that accelerates electrons trapped by the backward-SRS induced Langmuir wave. Finally, in Region III, when $n_e\gtrsim0.138n_c$, the two stage electron acceleration process by backward-SRS and forward-SRS occurs. 
	
	When $T_e=\SI{2.5}{keV}$, if $n_e>0.138n_c$ (Region III), then $p_{min}^F<p_m^B$, and electrons trapped by the backward-SRS induced Langmuir wave are further trapped and accelerated by the forward-SRS induced Langmuir wave. This results in the generation of suprathermal electrons with energy higher than $E_\phi^F$. As a result, one needs to reduce the electron density to be less than $n_e\sim0.138n_c$ in indirect-drive ICF experiments to avoid suprathermal electron generation with energy higher than $E_\phi^F$. From this work, the best choice of the gas fill for hohlraum targets in indirect-drive to suppress suprathermal electron generation is $0.108n_c\lesssim n_e\lesssim0.138n_c$ (Region II).

	It should be noted that the hot electron spectra and numbers via the anti-Stokes Langmuir decay instability cascade mechanism described here might be used as a method of inferring electron density information for the interior of hohlraum targets for future indirect-drive experiments, since the suprathermal electrons that are generated are very sensitive to the precise electron density (of course, this information is limited by the electric fields associated with space charge separation which will also need to be accounted for).
	
	Finally, the novel acceleration mechanism presented here explains the higher energy electrons generation in indirect-drive experiments. Based on these mechanisms discussed here, it can promote the future mitigation strategies of hot electrons for experiments on the NIF.
	And it also provides a promising approach to accelerate the electrons to higher energy as a hard X-ray source for radiography purposes. Indeed, the use of plasma optics and beam-combiners \cite{Kirkwood_2018NP,Kirkwood_2018POP} might aid this application. Future studies will also need to look at more precise requirements for the control of fast electron generation with the use of direct drive ICF, since suprathermal electron generation in the laser propagation direction during the compression phase is a critical parameter that must be controlled.

	\begin{acknowledgments}
		We would like to acknowledge useful discussions with C. Z. Xiao, L. Hao and Q. Wang. The authors gratefully acknowledge the support of the  ARCHER2 UK National Supercomputing Service, all of the staff of the Central Laser Facility and the Scientific Computing Department's SCARF supercomputing facility at the UKRI-STFC Rutherford Appleton Laboratory. 
		This research was supported by the Oxford-ShanghaiTech collaboration, the UKRI-EPSRC funded e674 ARCHER2 project and the National Natural Science Foundation of China (Grant Nos. 12005021, 11875091 and 11875093).
	\end{acknowledgments}

	\bibliography{SRS_manuscript}
	%\longbibliography{SRS_manuscript}
	
\end{document}